\documentclass[11pt, a4paper]{article}
\pdfoutput=1
\usepackage{jheppub}
\usepackage[latin1]{inputenc}
\usepackage{amsmath}
\usepackage{amsfonts}
\usepackage{amssymb}
\usepackage{latexsym}
\usepackage{mathrsfs}
\usepackage{graphicx}
\usepackage{color}
\usepackage{twistor}
\usepackage[all]{xy}

\title{Correlation functions, null polygonal Wilson loops, and local operators}

\author{Tim Adamo}

\affiliation{The Mathematical Institute \\ University of Oxford \\
	24-29 St.~Giles', Oxford OX1 3LB \\United Kingdom}
	
\emailAdd{adamo@maths.ox.ac.uk}

\abstract{We consider the ratio of the correlation function of $n+1$ local operators over the correlator of the first $n$ of these operators in planar $\cN=4$ super-Yang-Mills theory, and consider the limit where the first $n$ operators become pairwise null separated.  By studying the problem in twistor space, we prove that this is equivalent to the correlator of a $n$-cusp null polygonal Wilson loop with the remaining operator in general position, normalized by the expectation value of the Wilson loop itself, as recently conjectured by Alday, Buchbinder and Tseytlin.  Twistor methods also provide a BCFW-like recursion relation for such correlators.  Finally, we study the natural extension where $n$ operators become pairwise null separated with $k$ operators in general position.  As an example, we perform an analysis of the resulting correlator for $k=2$ and discuss some of the difficulties associated to fixing the correlator completely in the strong coupling regime.}

\begin{document}
\maketitle


\section{Introduction}
\label{Introduction}

There is now a considerable literature linking null polygonal Wilson loops to scattering amplitudes in the planar limit of $\cN=4$ super-Yang-Mills (SYM) theory.  Motivated by the AdS/CFT correspondence, Alday and Maldacena first conjectured the duality between the expectation value of a $n$ cusp null polygonal Wilson loop in the fundamental representation of the gauge group and $n$ particle gluon scattering amplitudes by studying the problem in the strong coupling regime (i.e., using string theory in the $AdS_{5}\times S^{5}$ geometry near the boundary) \cite{Alday:2007hr}.  Since then, a wide variety of studies have been performed at both strong and weak coupling which indicate that the duality should be true (c.f., \cite{Drummond:2007aua, Brandhuber:2007yx, Drummond:2007cf, Drummond:2007bm, Drummond:2008aq, Gorsky:2009dr}).  More recently, the advent of the twistor Wilson loop for $\cN=4$ SYM \cite{Mason:2010yk} has provided an efficient means of checking the duality for arbitrary NMHV degree and loop order at the level of the integrand (for both the Wilson loop and scattering amplitudes), and it has also been shown that the twistor Wilson loop has the same singularity structure as scattering amplitudes \cite{Bullimore:2011ni}.  This essentially constitutes a twistor-theoretic proof of Alday and Maldacena's original conjecture at the level of the integrand (see \cite{Adamo:2011pv} for a review).\footnote{There is some skepticism as to whether or not the loop integrand is a well-defined object; in the abscence of a meaningful regularization scheme, anomalies are present when the twistor calculations are translated to space-time (e.g., \cite{Belitsky:2011zm}).  Throughout this paper, we will adopt the philosophy that the loop integrand \emph{is} a meaningful object that could be regularized by some scheme which has not yet been fully realized.}

However, the duality between Wilson loops and other gauge-theoretic objects does not stop here.  In \cite{Eden:2010zz, Alday:2010zy, Eden:2010ce}, it was conjectured that, in the limit where the insertion points become pairwise null separated, the ratio of certain $n$-point correlation functions in $\cN=4$ SYM is equal to the expectation value of a null polygonal Wilson loop in the adjoint representation.  More formally, if $\{\cO(x_{i})\}_{i=1,\ldots,n}$ are gauge invariant local operators in $\cN=4$ SYM, then this conjecture takes the form:
\be{corrW}
\lim_{x_{i,i+1}^{2}\rightarrow 0}\frac{\la \cO(x_{1})\ldots\cO(x_{n})\ra}{\la\cO(x_{1})\ldots\cO(x_{n})\ra^{\mathrm{tree}}}=\la W^{n}_{\mathrm{adj}}[C]\ra \xrightarrow{\mathrm{planar}\:\mathrm{limit}} \la W^{n}[C]\ra^{2},
\ee
where $C$ is the resulting null polygon with $n$ cusps, $W^{n}_{\mathrm{adj}}$ is the Wilson loop in the adjoint representation, and $W^{n}$ the Wilson loop in the standard fundamental representation.  Difficult calculations on space-time have confirmed this conjecture through examples \cite{Eden:2011yp, Eden:2011ku}, but it can be proven efficiently in twistor space at the level of the integrand \cite{Adamo:2011dq}.

In this paper, we consider yet another Wilson loop duality- in particular, the conjecture of \cite{Alday:2011ga}, which extends \eqref{corrW} by including an additional operator in general position (i.e., not null separated from any of the other insertion points).  In the planar limit, Alday, Buchbinder and Tseytlin conjecture that
\be{locW1}
\lim_{x_{i,i+1}^{2}\rightarrow 0}\frac{\la \cO(x_{1})\ldots\cO(x_{n})\cO(y)\ra}{\la\cO(x_{1})\ldots\cO(x_{n})\ra}\sim \frac{\la W^{n}[C]\cO(y)\ra}{\la W^{n}[C]\ra}\equiv \cC^{n}_{1}(W^{n}, y),
\ee
based upon an integration-by-parts argument in the path integral when $\cO(y)=\cO_{\mathrm{dil}}(y)$, the dilaton operator.  The motivations for considering such an object are many: these mixed correlators are a natural candidate for interpolating between Wilson loops and generic correlation functions; their structure is highly constrained by conformal invariance; studying $\cC^{n}_{1}$ provides information about the Wilson loop OPE \cite{Berenstein:1998ij}; and while $\la W^{n}[C]\ra$ is known to be UV divergent due to the $n$ cusps of $C$, the ratio on the right side of \eqref{locW1} appears to be finite \cite{Alday:2011ga}.

Recent work at weak coupling has confirmed this conjecture for twist-2 local operators using dimensional regularization \cite{Engelund:2011fg}; our main tasks will be to prove \eqref{locW1} for \emph{any} (gauge-invariant) local operator at the level of the loop integrand.  The key to the proof will be to translate the problem into the language of twistor theory, whereby the result follows relatively easily using the methods of \cite{Adamo:2011dq}.  Additionally, we use these twistor methods to derive a BCFW-like recursion formula for the correlator on the right-hand side of \eqref{locW1} and can also prove natural generalizations of the conjecture involving additonal operator insertions and null limits, as suggested by \cite{Engelund:2011fg}.  The key advantage of the twistor approach is its ability to elegantly (and efficiently) handle the supersymmetry of $\cN=4$ SYM while simultaneously providing a geometric description of the null limits under consideration.

In section \ref{twistor}, we provide a brief review of the basic aspects of twistor theory and how they are used to translate the problem of correlation functions into a twistorial framework.  Section \ref{Proof} begins by providing the intuition behind \eqref{locW1}, and then proceeds directly with a proof based on the twistor-theoretical tools developed in the previous discussion, with many of the explicit calculations relegated to appendix \ref{calcs}.  We then use the twistor framework to derive a BCFW-type recursion for the Wilson loop - local operator correlator in section \ref{BCFW}.  Our methods also apply to a straightforward generalization of \eqref{locW1}, which includes the correlator of a null polygonal Wilson loop with any number of local operators in general position (i.e., not null separated from each other or the cusps of the Wilson loop), which we discuss in section \ref{strong}.  As an example, we include a brief exploration of this new generalization by studying an ansatz for the functional form of the correlator in terms of conformal cross-ratios, and performing some simple strong coupling calculations when $n=4$ and there are two operator insertions.  Section \ref{conclusion} concludes.


\section{Twistor Theory Background}
\label{twistor}

Since our main tool for proving the conjecture of \cite{Alday:2011ga} will be twistor theory, it would do to have a brief review of the necessary machinery before proceeding.  More detailed overviews of classical twistor theory can be found in \cite{Huggett:1985, Penrose:1986}; a review more adapted to the problem at hand can be found in \cite{Adamo:2011pv}.

\subsection{Twistor basics}

For us, twistor space $\PT$ is a suitable open neighborhood of the complex projective supermanifold $\CP^{3|4}$.  We work with homogeneous coordinates
\begin{equation*}
Z^{I}=(\lambda_{A}, \mu^{A'}, \chi^{a})=(Z^{\alpha},\chi^{a}),
\end{equation*}
where $\lambda_{A}$ and $\mu^{A'}$ are 2-component Weyl spinors and $\chi^{a}$, $a=1,\ldots,4$ are anti-commuting Grassmann variables indexing the $\SU(4)_{R}$ R-symmetry of $\cN=4$ SYM.  These twistor coordinates are related to the coordinates $(x^{AA'}, \theta^{Aa})$ on chiral super-Minkowski space $\M$ by the standard incidence relations:
\be{incidence}
\mu^{A'}=i x^{AA'}\lambda_{A}, \qquad \chi^{a}=\theta^{Aa}\lambda_{A}.
\ee
Hence, we see that a point $(x,\theta)\in\M$ corresponds to line $X\cong\CP^{1}$ in $\PT$ defined by these equations.  As $\CP^{3|4}$ is a Calabi-Yau supermanifold, $\PT$ is equipped with a canonical global holomorphic measure (or global section of the Berezinian sheaf) which we denote
\begin{equation*}
\D^{3|4}Z=\epsilon_{\alpha\beta\gamma\delta}Z^{\alpha}\d Z^{\beta}\d Z^{\gamma}\d Z^{\delta} \d^{4}\chi.
\end{equation*}

One of the most important tools in twistor theory is the Penrose transform, which allows us to represent zero-rest-mass (z.r.m.) fields on space-time in terms of cohomological data on twistor space.  If $U\subset\CP^{3}$ is a (suitably chosen) open subset of the bosonic reduction of twistor space, then the Penrose transform is manifested by the following isomorphism:
\begin{equation*}
H^{1}(U, \cO(2h-2))\cong\left\{\mbox{On-shell z.r.m. fields on $\M$ of helicity }h\right\},
\end{equation*}
where $\cO(n)$ is the sheaf of holomorphic functions which are homogeneous of degree $n$.  For us, the main upshot of this is that it allows us to encode the entire $\cN=4$ SYM multiplet into a single homogeneous superfield on $\PT$:
\be{tfield}
\cA = a +\chi^{a}\tilde{\psi}_{a}+\frac{\chi^{a}\chi^{b}}{2!}\phi_{ab}+\frac{\epsilon_{abcd}}{3!}\chi^{a}\chi^{b}\chi^{c}\psi^{d}+\frac{\chi^{4}}{4!}g,
\ee
which has no components in the anti-holomorphic fermionic directions, and each bosonic coefficient corresponds to a space-time field via the Penrose transform \cite{Witten:2003nn}.  This transform can be realized quite explicitly; for example, the scalars are given by:
\begin{equation*}
\Phi_{ab}(x)=\int_{X}\la\lambda\d\lambda\ra \wedge \phi_{ab}(Z)|_{X},
\end{equation*}
where the restriction of $\phi_{ab}$ to $X\cong\CP^{1}$ is given by the incidence relations \eqref{incidence}, and $\la\lambda\d\lambda\ra=\lambda_{A}\d\lambda^{A}$ is the weight $+2$ holomorphic measure on $X$.  

Let us assume that we work with a gauge group $G=\SU(N)$ whose complexified Lie algebra is $\mathfrak{g}^{\C}=\mathfrak{sl}_{N}$.  The superfield $\cA$ provides an \emph{off-shell}\footnote{By off-shell, we mean that we do not \textit{a priori} impose the condition $\dbar\cA=0$.} description of the $\cN=4$ supermultiplet in twistor-space, and acts as a $(0,1)$-connection on a gauge bundle $E\rightarrow\PT$, which has $c_{1}(E)=0$, $\End(E)\cong\mathfrak{sl}_{N}$, and $\cA\in\Omega^{0,1}(\PT,\cO\otimes\End(E))$.  This serves as the main variable in the twistor action formulation of $\cN=4$ SYM \cite{Mason:2005zm, Boels:2006ir}:
\be{TA}
S[\cA]=\frac{i}{2\pi}\int_{\PT}\D^{3|4}Z\wedge\tr\left(\cA\wedge\dbar\cA+\frac{2}{3}\cA\wedge\cA\wedge\cA\right)+\lambda \int_{\M}\d^{4|8}X\log\det\left(\dbar+\cA\right)|_{X},
\ee
where $\lambda$ is the 't Hooft coupling, $\d^{4|8}X$ is the measure over the space of $X\cong\CP^{1}$ in $\PT$, and $(\dbar+\cA)|_{X}$ is the complex structure induced by $\cA$ restricted to the line $X$.  The holomorphic Chern-Simons term in $S[\cA]$ accounts for the self-dual sector of the Yang-Mills theory by the supersymmetric Ward correspondence, while the second non-local term generates the ASD interactions.  This can be seen by perturbatively expanding the $\log\det(\dbar+\cA)$ in the second term, which generates the MHV vertices of the theory.  

This twistor action has significantly more gauge freedom than the space-time Yang-Mills action,
\begin{equation*}
(\dbar+\cA)\rightarrow \gamma(\dbar+\cA)\gamma^{-1}, \qquad \gamma\in \Gamma(E, \End(E)),   
\end{equation*}
and this can be reduced to that of space-time gauge transformations by imposing the condition that $\cA$ be holomorphic upon restriction to the $\CP^1$ fibers of twistor space \cite{Woodhouse:1985id}.  An alternative gauge fixing is provided by the so-called `CSW gauge', which gives a foliation of twistor space induced by a fixed reference twistor $Z_{*}$ and demands that $\cA$ vanish upon restriction to the leaves of this foliation.  More explicitly, we have:
\begin{equation*}
\overline{Z^{I}_{*}\frac{\partial}{\partial Z^{I}}}\lrcorner\cA =0.
\end{equation*}  
It has been shown that the Feynman rules of the twistor action in the CSW gauge reproduce the MHV formalism of \cite{Cachazo:2004kj} on twistor space \cite{Adamo:2011cb}.  For the remainder of this paper, while working with the twistor action we assume that CSW gauge has been imposed; this means that the twistor space propagator takes the form:
\be{tprop}
\Delta_{*}(Z,Z')^{i\;k}_{j\;l}=\bar{\delta}^{2|4}(Z, *, Z')\left(\delta^{i}_{l}\delta^{k}_{j}-\frac{1}{N}\delta^{i}_{j}\delta^{k}_{l}\right),
\ee
where $i,j,k$, and $l$ are gauge indices, $N$ is the rank of the gauge group, and $\bar{\delta}^{2|4}$ is a $(0,2)$-current on $\PT$
\begin{equation*}
\bar{\delta}^{2|4}(Z,*,Z')=\int_{\C^{2}}\frac{\d s}{s}\frac{\d t}{t}\bar{\delta}^{4|4}(Z+s Z_{*}+t Z'),
\end{equation*}
which enforces the (projective) collinearity of its three arguments.

\subsection{Local operators and Wilson loops in twistor space}

The correlation functions we are interested in involve local operators in our $\cN=4$ gauge theory.  These could include the Konishi or dilaton operators (or indeed any chiral primary operators), but for now we will restrict our attention to the `1/2-BPS' operators:\footnote{Since we work at the level of the loop integrand, a correlation function of any local operators will simply be a rational function.  However, 1/2-BPS operators do not require renormalization, and are thus the best objects to consider if we want our claims to extend to the full loop \emph{integral}.}
\be{BPS}
\cO(x)=\cO_{abcd}(x)=\tr(\Phi_{ab}(x)\Phi_{cd}(x))-\frac{\epsilon_{abcd}}{12}\tr(\Phi^{2}(x)).
\ee
For an abelian gauge group, it is easy to see how to express $\cO$ in twistor space using the Penrose transform:
\begin{multline*}
\cO^{\U(1)}(x)=\int_{X\times X}\la\lambda\d\lambda\ra\wedge\la\lambda'\d\lambda'\ra\wedge\phi_{ab}(\lambda)\wedge\phi_{cd}(\lambda') \\
-\frac{\epsilon_{abcd}}{12}\int_{X\times X}\la\lambda\d\lambda\ra\wedge\la\lambda'\d\lambda'\ra\wedge\phi^{ef}(\lambda)\wedge\phi_{ef}(\lambda'),
\end{multline*}
where $\phi_{ab}(\lambda)$ denotes the pullback of $\phi_{ab}$ to the line $X$ charted by $\lambda$.  A natural supersymmetric generalization, which we shall use from now on, is given by taking $\frac{\partial^{2}\cA}{\partial\chi^{2}}$ instead of $\phi_{ab}$:
\begin{multline}\label{abelian}
\cO^{\U(1)}(x,\theta)=\int_{X\times X}\la\lambda\d\lambda\ra\wedge\la\lambda'\d\lambda'\ra\wedge\frac{\partial^{2}\cA}{\partial\chi^{a}\partial\chi^{b}}(\lambda)\wedge\frac{\partial^{2}\cA}{\partial\chi^{c}\partial\chi^{d}}(\lambda') \\
-\frac{\epsilon_{abcd}}{12}\int_{X\times X}\la\lambda\d\lambda\ra\wedge\la\lambda'\d\lambda'\ra\wedge\frac{\partial^{2}\cA}{\partial\chi_{e}\partial\chi_{f}}(\lambda)\wedge\frac{\partial^{2}\cA}{\partial\chi^{e}\partial\chi^{f}}(\lambda'),
\end{multline}
where
\begin{equation*}
\frac{\partial^{2}\cA}{\partial\chi^{a}\partial\chi^{b}}=\phi_{ab}+\epsilon_{abcd}\chi^{c}\psi^{d}+\frac{1}{2!}\epsilon_{abcd}\chi^{c}\chi^{d}g.
\end{equation*}

\begin{figure}
\centering
\includegraphics[width=1.7 in, height=1 in]{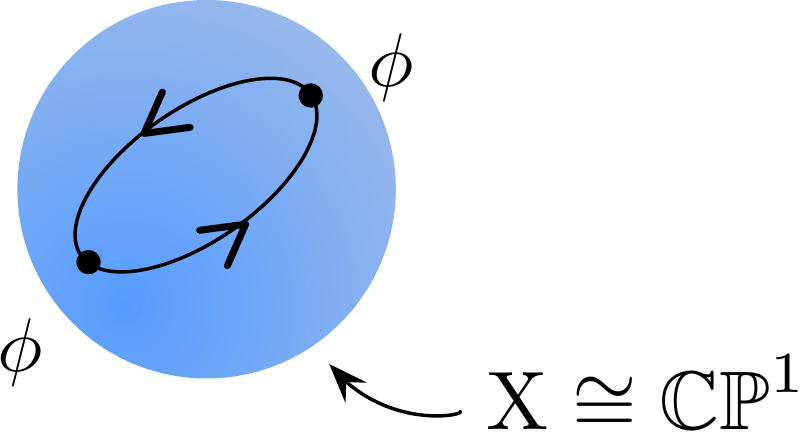}\caption{\textit{The twistor space form of the local space-time operator} $\tr\Phi^2(x)$, \textit{involving holomorphic Wilson lines; arrows indicate the flow of the color trace.}}\label{operator}
\end{figure}

Unfortunately, for a non-abelian gauge group \eqref{abelian} cannot suffice since there is no way for us to compare fibers of the gauge bundle $E$ at different points on $X$; this requires a frame for $E\rightarrow\PT$ which provides a holomorphic trivialization of $E|_{X}$.  For dimensional reasons, $E$ is holomorphic upon restriction to $X$ and we have assumed that $E$ is topologically trivial, so all that is required is a gauge transformation $\gamma$ which obeys:
\begin{equation*}
\gamma(\dbar+\cA)|_{X}\gamma^{-1}=\dbar|_{X}.
\end{equation*}
As it turns out, such a $\gamma$ can be found generically, since $X$ is rational and linearly embedded.  If we define
\be{frame}
U_{X}(\lambda,\lambda')=\gamma(x,\lambda)\gamma^{-1}(x,\lambda'),
\ee
then $U_{X}$ is formally a Green's function for $(\dbar+\cA)|_{X}$, and acts as
\begin{equation*}
U_{X}(\lambda,\lambda)=\mathbb{I}, \qquad U_{X}(\lambda,\lambda'): E|_{\lambda'}\rightarrow E|_{\lambda}.
\end{equation*}
Thus, it is natural to interpret $U_{X}$ as the twistor space parallel propagator for the gauge bundle $E$ along $X$.  This allows us to write down an immediate non-abelian generalization of \eqref{abelian} for our 1/2-BPS operators \cite{Adamo:2011dq}:
\begin{multline}\label{nabelian}
\cO(x,\theta)=\int_{X\times X}\la\lambda\d\lambda\ra \la\lambda' \d\lambda'\ra \tr\left[U_{X}(\lambda,\lambda')\frac{\partial^{2}\cA(\lambda')}{\partial\chi^{a}\partial\chi^{b}}U_{X}(\lambda',\lambda)\frac{\partial^{2}\cA(\lambda)}{\partial\chi^{c}\partial{\chi}^{d}}\right] \\
-\frac{\epsilon_{abcd}}{12}\int_{X\times X}\la\lambda\d\lambda\ra \la\lambda' \d\lambda'\ra \tr\left[U_{X}(\lambda,\lambda')\frac{\partial^{2}\cA(\lambda')}{\partial\chi_{e}\partial{\chi}_{f}}U_{X}(\lambda',\lambda)\frac{\partial^{2}\cA(\lambda)}{\partial\chi^{e}\partial{\chi}^{f}}\right].
\end{multline}
The bosonic portion of this operator is depicted in figure \ref{operator}.  Since we use the fully supersymmetric parallel propagator $U_{X}$, $\cO(x,\theta)$ corresponds to the chiral part of the 1/2-BPS multiplet.

The parallel propagator $U_{X}$ also allows us to define the twistor Wilson loop of \cite{Mason:2010yk, Bullimore:2011ni}.  Consider a null polygon $C$ in $\M$ with $n$ cusps labelled by $(x_{i},\theta_{i})$.  The Wilson loop $W^{n}[C]$ is defined to be the holonomy of the space-time gauge connection about the null polygon $C$.  But we can lift this to twistor space, where the null polygon is replaced by a chain of intersecting $\CP^1$s.  In particular, the cusp $x_{i}$ becomes $X_{i}\cong\CP^1$, and intersects $X_{i-1}$ and $X_{i+1}$ in points $Z_{i-1}$ and $Z_{i}$ respectively.  The space-time gauge connection becomes the $(0,1)$-connection $\cA$ on $E\rightarrow\PT$, and its holonomy is the trace of the parallel propagation of this connection around the resulting nodal curve in twistor space.  Hence, we define
\be{WL}
W^{n}[C]=\tr\left[ U_{X_n}(\lambda_{n},\lambda_{n-1})U_{X_{n-1}}(\lambda_{n-1},\lambda_{n-2})\cdots U_{X_1}(\lambda_{1},\lambda_{n})\right],
\ee
to be the twistorial representation of the Wilson loop around $C$ in the fundamental representation of the gauge group.  It has now been confirmed that \eqref{WL} coincides with the supersymmetric space-time Wilson loop of Caron-Huot \cite{CaronHuot:2010ek} up to terms proportional to the equations of motion \cite{Belitsky:2011zm}.

From now on, we will abuse notation and denote coordinates on $\M$ by their bosonic part only; in other words, $(x,\theta)$ will be abbreviated to $x$.  Additionally, we make no distinction between the null polygon in space-time and the corresponding nodal elliptic curve in twistor space, denoting both by $C$.

With these tools, we set out to prove the conjecture of \eqref{locW1} and generalize it via twistor methods.


\section{Proving the Correspondence}
\label{Proof}

\subsection{Some motivation: the dilaton operator}
\label{dilmot}

Before embarking on our twistor theoretic proof of \eqref{locW1}, we first provide a bit of motivation for why one could expect such a correspondence to hold.  Consider the correlator of $n$ of our 1/2-BPS operators \eqref{BPS} with the dilaton operator, which for a $\SU(N)$ gauge theory takes the form \cite{Liu:1999kg}:
\be{dilaton}
\cO_{\mathrm{dil}}(y)=\hat{c}_{\mathrm{dil}}\tr\left(F^{2}+\Phi^{ab}\partial^{AA'}\partial_{AA'}\Phi_{ab}+\tilde{\Psi}_{a}^{A'}\partial_{AA'}\Psi^{a\;A}+\cdots \right), \qquad \hat{c}_{\mathrm{dil}}=\frac{\pi^2}{3\sqrt{3}N},
\ee
with the dots representing terms of higher order in the gauge coupling.  In essence, this means that (following a proper re-scaling) the dilaton operator is just the $\cN=4$ SYM Lagrangian.  Now consider
\begin{equation*}
\lim_{x_{i,i+1}^{2}\rightarrow 0}\left\la \cO(x_{1})\cdots\cO(x_{n})\int_{\M}\d^{4}y \cO_{\mathrm{dil}}(y)\right\ra,
\end{equation*}
where we adopt the usual notation
\begin{equation*}
x^{\mu}_{i,j}\equiv x^{\mu}_{i}-x^{\mu}_{j}.
\end{equation*}

Working in the planar limit (i.e, $N\rightarrow\infty$) and taking for granted the Wilson loop / correlation functions correspondence given by \eqref{corrW}, we can work inside the path integral to find \cite{Alday:2011ga}:
\begin{multline*}
\lim_{x_{i,i+1}^{2}\rightarrow 0}\left\la \cO(x_{1})\cdots\cO(x_{n})\int_{\M}\d^{4}y \;\cO_{\mathrm{dil}}(y)\right\ra \sim \left\la (W^{n}[C])^{2} \int_{\M}\d^{4}y\; \cO_{\mathrm{dil}}(y)\right\ra \\
= \int [\mathcal{D}\mathcal{F}] (W^{n}[C])^{2} \int_{\M}\d^{4}y \cO_{\mathrm{dil}}(y) \:\exp\left[- \lambda\int_{\M}\d^{4}y \cO_{\mathrm{dil}}(y)\right] \\
= - \int [\mathcal{D}\mathcal{F}] (W^{n}[C])^{2} \frac{\partial }{\partial \lambda}e^{-\lambda S[\mathcal{F}]} = \int [\mathcal{D}\mathcal{F}] \frac{\partial (W^{n}[C])^{2}}{\partial \lambda} e^{-\lambda S[\mathcal{F}]} \\
= 2 \la W^{n}[C] \ra \left\la W^{n}[C] \int_{\M}\d^{4}y \cO_{\mathrm{dil}}(y)\right\ra,
\end{multline*}
where $\lambda$ is the 't Hooft coupling, and the third line follows by integration by parts within the path integral.  If we assume that this argument still works without integrating over the position of the dilaton operator, then it appears to justify the statement
\begin{equation*}
\lim_{x_{i,i+1}^{2}\rightarrow 0}\frac{\la \cO(x_{1})\cdots\cO(x_{n})\cO_{\mathrm{dil}}(y)\ra}{\la \cO(x_{1})\cdots\cO(x_{n})\ra}\sim \frac{\la W^{n}[C]\cO_{\mathrm{dil}}(y)\ra}{\la W^{n}[C]\ra}.
\end{equation*}

However, there is no immediately obvious reason why this should still work when the integral over the position of the dilaton insertion has been omitted.  Furthermore, we would like to be able to prove \eqref{locW1} for \emph{any} local operators, not just the dilaton operator.  To this end, we now apply our twistor machinery laid out in \S \ref{twistor}.

\subsection{Proof via twistor theory}

Without loss of generality, let all operators in question be those 1/2-BPS operators discussed in \S \ref{twistor}. We can easily adapt the following to include Konishi or dilaton operators (whose correlation functions are perfectly well-defined at the level of the integrand), but prefer to deal with the gauge invariant 1/2-BPS operators since we do not discuss renormalization of the loop integrals.  Now, provided all limits exist (as we will show), we can separate the limit of interest as
\be{eqn: l2}
\lim_{x^{2}_{i,i+1}\rightarrow 0}\frac{\la \cO(x_{1})\cdots\cO(x_{n})\cO(y)\ra}{\la \cO(x_{1})\cdots\cO(x_{n})\ra^{\mathrm{tree}}}\times \lim_{x^{2}_{i,i+1}\rightarrow 0}\frac{\la \cO(x_{1})\cdots\cO(x_{n})\ra^{\mathrm{tree}}}{\la \cO(x_{1})\cdots\cO(x_{n})\ra},
\ee
where the insertion $y$ is in general position (i.e., not null separated from any of the $x_{i}$).

However, using the results of \cite{Adamo:2011dq} on null limits of correlation functions expressed in \eqref{corrW}, it is easy to see that the limit we are actually interested in calculating is computed by:
\be{newlimit}
\lim_{x^{2}_{i,i+1}\rightarrow 0}\frac{\la \cO(x_{1})\cdots\cO(x_{n})\cO(y)\ra}{\la \cO(x_{1})\cdots\cO(x_{n})\ra^{\mathrm{tree}}}\times \frac{1}{\la W^{n}_{\mathrm{adj}}[C]\ra}.
\ee
It is well known that the tree level contribution in the denominator goes as
\begin{equation*}
\la \cO(x_{1})\cdots\cO(x_{n})\ra^{\mathrm{tree}} \sim \frac{1}{x_{12}^{2}x_{23}^{2}\cdots x_{n1}^{2}},
\end{equation*}
so as we evaluate the expectation value in the numerator, we can neglect all those contributions which do not have a divergence in the null limit to counterbalance this classical factor.  We perform this calculation by working twistorially, using the operators \eqref{nabelian} and the twistor action for $\cN=4$ SYM \eqref{TA}.

\begin{figure}
\centering
\includegraphics[width=3 in, height=1.5 in]{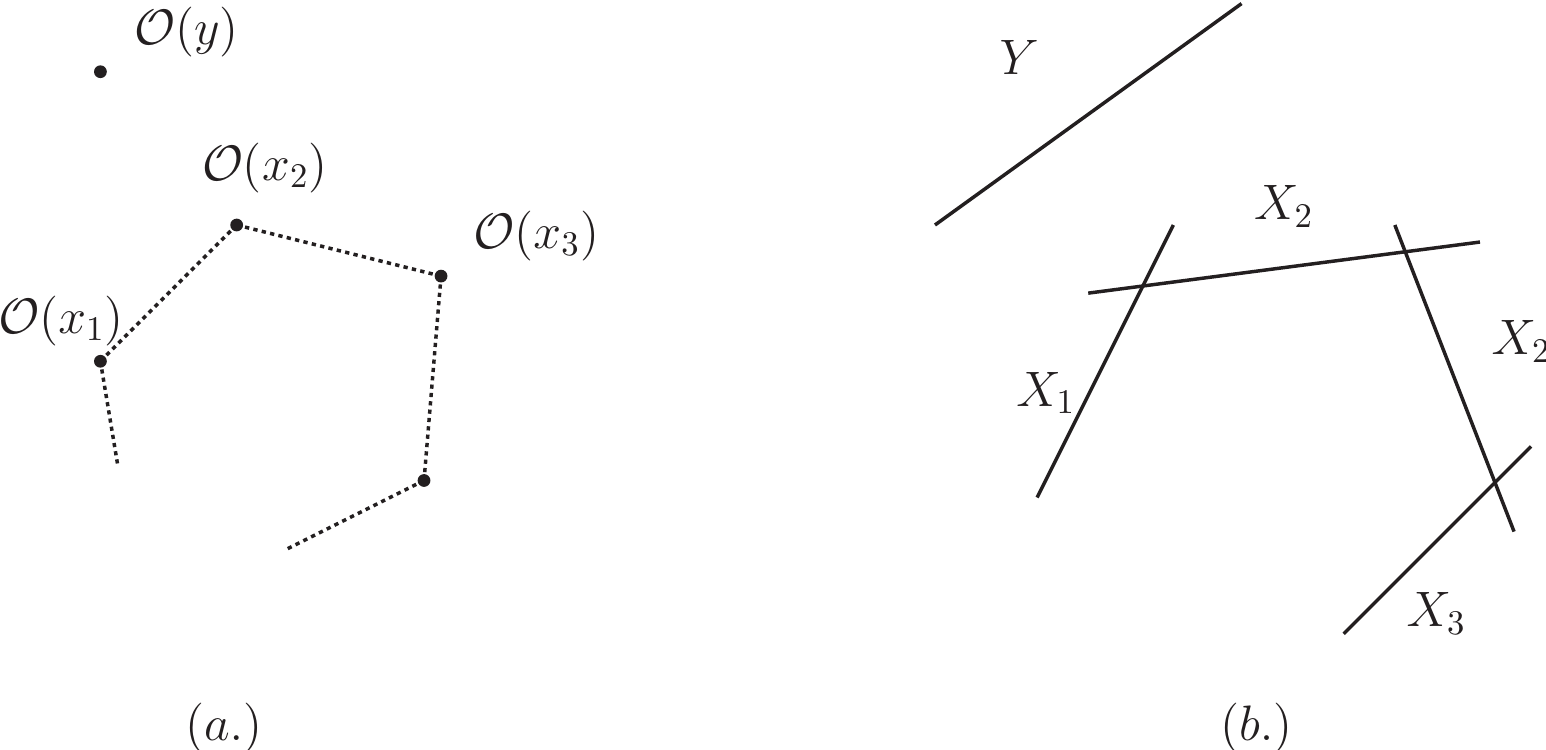}\caption{\textit{The geometry of the null limit in} (a.) \textit{space-time, and} (b.) \textit{twistor space.}}\label{figure 1}
\end{figure}
Geometrically, this limit is manifested rather nicely in twistor space.  We begin with $n+1$ lines $X_{1},\ldots,X_{n},Y\subset\PT$ on which each of our operators `lives.'  In the limit, the first $n$ of these intersect each other sequentially to form the nodal curve corresponding to the resulting null polygon in $\M$; the final operator in general position, $\cO(y)$, lies on a line $Y$ which does not intersect \emph{any} of the others.  This configuration is illustrated in figure \ref{figure 1}.

We now  evaluate the numerator of \eqref{newlimit} by applying Wick's theorem to the twistorial path integral
\begin{equation*}
\int [\D\cA] \cO(x_{1})\cdots\cO(x_{n})\cO(y)e^{-S[\cA]},
\end{equation*}
under the assumptions of \emph{normal ordering} and \emph{genericity}.  The normal ordering assumption means that we can exclude any contractions between fields or frames inserted on the same lines in twistor space; the genericity assumption means that the MHV vertices generated by the second term in the twistor action \eqref{TA} are not null separated from any of the operator insertions.  Twistorially, this latter condition is equivalent to saying that the lines generated by a perturbative expansion of the $\log$-$\det$ in the twistor action do not intersect any of the lines where operator insertions live.

Hence, we are left with the following contractions to consider:
\begin{itemize}
\item Contractions between an operator and a MHV vertex from the twistor action

\item Contractions between operators and frames on the $\{X_{i}\}_{i=1,\ldots,n}$

\item Contractions between operators and frames on $\{X_{i}\}$ and $Y$
\end{itemize}

These contractions can be performed explicitly using the twistor propagator \eqref{tprop} as in \cite{Adamo:2011dq}.  In appendix \ref{calcs} we show that most of these contractions are finite or vanishing in the null limit; for instance, the genericity assumption assures that all contractions of the first type are finite in the null limit, and can therefore be disregarded.  Similarly, all contractions amongst the operators and frames of the $\{X_i\}$ are finite or vanishing except for those between $\frac{\partial^{2}\cA}{\partial\chi^{2}}$ on adjacent lines (i.e., between $X_{i}$ and $X_{i+1}$).  These contractions produce a factor of 
\begin{equation*}
\int_{X_{i}\times X_{i+1}}\la\lambda_{i}\d\lambda_{i}\ra\la\lambda_{i+1}\d\lambda_{i+1}\ra \left\la \overbrace{\frac{\partial^{2}\cA}{\partial\chi^{a}\partial\chi^{b}}|_{X_{i}}\frac{\partial^{2}\cA}{\partial\chi^{c}\partial\chi^{d}}|_{X_{i+1}}} \right\ra = \frac{\epsilon_{abcd}}{(x_{i}-x_{i+1})^2},
\end{equation*}
which exactly counterbalances the tree-level denominator of \eqref{newlimit}!  Furthermore, these contractions leave two holomorphic frames $U_{X}(\lambda,\lambda')$ on each of the $X_{i}$, and in the null limit, the trace around these (now intersecting) lines yields the integrand of the twistor Wilson loop in the adjoint representation.  As we have thus far ignored any contractions involving the generically placed operator $\cO(y)$, we are left with
\be{nplim1}
\lim_{x^{2}_{i,i+1}\rightarrow 0}\frac{\la \cO(x_{1})\cdots\cO(x_{n})\cO(y)\ra}{\la \cO(x_{1})\cdots\cO(x_{n})\ra^{\mathrm{tree}}}=\la W^{n}_{\mathrm{adj}}[C]\cO(y) \ra.
\ee

\begin{figure}
\centering
\includegraphics[width=4 in, height=1.5 in]{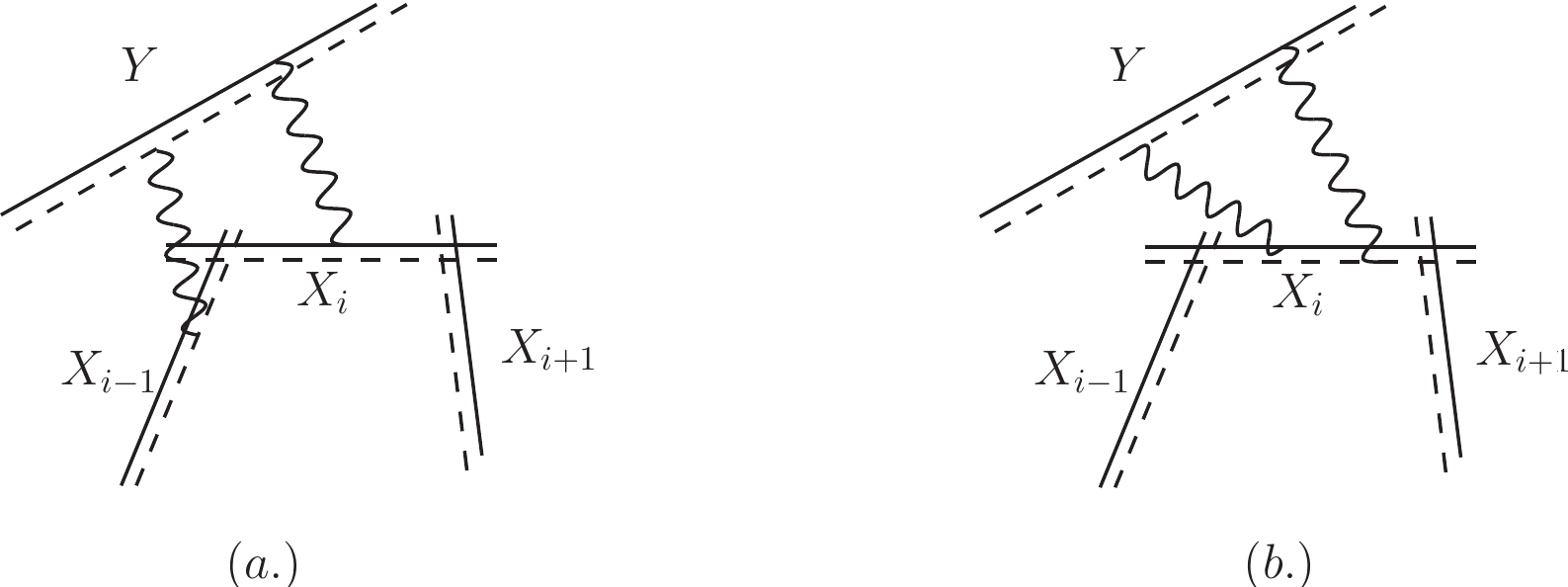}\caption{\textit{Contractions which are} (a.) \textit{leading, and} (b.) \textit{suppressed in the planar limit.  The solid and dashed lines are meant to distinguish the fundamental and anti-fundamental representations.}}\label{figure 2}
\end{figure}

Now, passing to the planar limit of the gauge theory (i.e., $N\rightarrow\infty$), the twistor propagator \eqref{tprop} suppresses any mixing between fundamental and anti-fundamental gauge indices due to the second term:
\begin{equation*}
\Delta_{*}(Z,Z')^{i\;k}_{j\;l}=\bar{\delta}^{2|4}(Z, *, Z')\left(\delta^{i}_{l}\delta^{k}_{j}-\frac{1}{N}\delta^{i}_{j}\delta^{k}_{l}\right).
\end{equation*}
So in the planar limit, we can decompose the adjoint representation into the product of fundamental and anti-fundamental representations at the level of the Wilson loop to write
\begin{equation*}
\la W^{n}_{\mathrm{adj}}[C]\cO(y) \ra = \la W^{n}[C] \widetilde{W}^{n}[C]\cO(y) \ra.
\end{equation*}
The final step is to factor out an expectation value of one of these Wilson loops.  Once again, this is made possible by the planar limit, which suppresses contractions between operators and frames on $Y$ with the Wilson loops which mix fundamental and anti-fundamental representations.  Figure \ref{figure 2} illustrates the distinction between the contractions which are suppressed and those which survive in the planar limit; a double line notation is adopted to distinguish between the fundamental and anti-fundamental representations.

This means that in the large $N$ limit, we have
\be{planar}
\la W^{n}[C] \widetilde{W}^{n}[C]\cO(y) \ra = 2 \la W^{n}[C] \ra \la W^{n}[C] \cO(y)\ra,
\ee
where we have dropped the irrelevant representation notation.  This is a rather explicit manifestation of the following general heuristic for a planar gauge theory: given three operators $\cO_{1}$, $\cO_{2}$, and $\cO_{3}$, their expectation value should obey
\begin{equation*}
\la \cO_{1} \cO_{2} \cO_{3} \ra = \la \cO_{1}\ra \la \cO_{2} \cO_{3}\ra + \la \cO_{2}\ra \la \cO_{3} \cO_{1}\ra + \la \cO_{3}\ra \la \cO_{1} \cO_{2}\ra.
\end{equation*}
In the case of \eqref{planar}, two of these terms are equal while the third vanishes due to normal ordering.

Equations \eqref{nplim1} and \eqref{planar} amount to a proof of the conjecture stated in \eqref{locW1}, as desired.  More formally, this can be stated as:

\begin{propn}\label{locP1}
Let $\{\cO(x_{i}), \cO(y)\}_{i=1,\ldots,n}$ be gauge invariant local operators in $\cN=4$ SYM, and $C$ be the null polygon resulting from the limit where the first $n$ of these operators become pairwise null separated (i.e., $x_{i,i+1}^{2}=0$).  Then at the level of the integrand,
\be{locW}
\lim_{x^{2}_{i,i+1}\rightarrow 0}\frac{\la \cO(x_{1})\cdots\cO(x_{n})\cO(y)\ra}{\la \cO(x_{1})\cdots\cO(x_{n})\ra} = \frac{\la W^{n}_{\mathrm{adj}}[C]\cO(y)\ra}{\la W^{n}_{\mathrm{adj}}[C]\ra} \xrightarrow{\mathrm{planar}\:\mathrm{limit}} 2\frac{\la W^{n}[C]\cO(y)\ra}{\la W^{n}[C]\ra},
\ee
where all expectation values are assumed to be generic and normal ordered, and $W^{n}[C]$ is the Wilson loop in the fundamental representation.
\end{propn}


\section{BCFW-like Recursion Relations}
\label{BCFW}

In this section, we will show that by performing a one-parameter deformation of the null polygonal Wilson loop, we can derive a BCFW-like recursion relation for the correlator
\begin{equation*}
\la W^{n}[C]\cO(y)\ra.
\end{equation*}
As in the previous section, the key to doing this will be studying the problem in twistor space.  

\begin{figure}
\centering
\includegraphics[width=3.5 in, height=2.5 in]{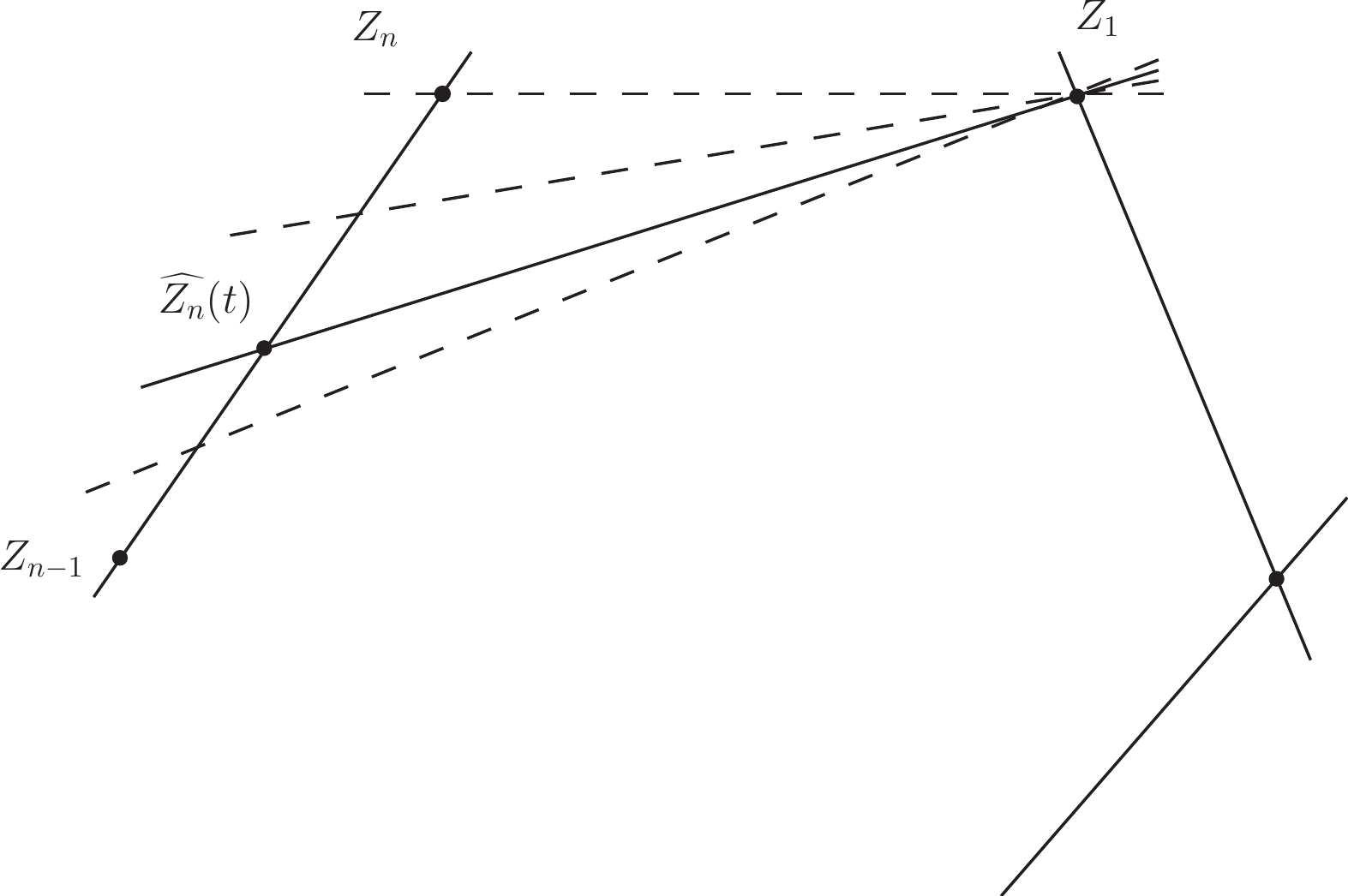}\caption{\textit{The BCFW-like deformation at the level of the twistor Wilson loop.}}\label{BCF1}
\end{figure}

Recall from \eqref{WL} the definition of the null polygonal Wilson loop in twistor space:
\begin{multline}\label{WL*}
W^{n}[C]\equiv W[1,2,\ldots, n]=\tr\: \mathrm{Hol}_{Z_n}[C]= \\
\tr\left[ U(Z_{n},Z_{n-1})U(Z_{n-1},Z_{n-2})\cdots U(Z_{1},Z_{n})\right],
\end{multline}
where $\mathrm{Hol}_{Z}[C]$ denotes the holonomy about $C$ at base point $Z$ and the $Z_{i}$ are the nodes of the resulting curve in twistor space.  With an abuse of notation, we will refer to this nodal curve in twistor space as $C$ also.  A BCFW-like deformation of the twistor Wilson loop is captured by performing a one-parameter shift of one of these nodes.  Without loss of generality, we can take
\be{BCFW1}
\widehat{Z_{n}}(t)=Z_{n}+tZ_{n-1}, \qquad t\in\C,
\ee
which shifts the $n^{\mathrm{th}}$ node along the line $(n-1, n)\cong\CP^{1}\subset\PT$, as illustrated in figure \ref{BCF1}.  Of course, \eqref{BCFW1} results in a one-parameter family of nodal curves in twistor space and their corresponding family of Wilson loops:
\begin{equation*}
C(t)=(1,2)\cup (2,3)\cup\cdots \cup (n-1, \hat{n}(t))\cup (\hat{n}(t), 1), \qquad W[C(t)]=W[1,\ldots n-1, \hat{n}(t)],
\end{equation*}
where we have adopted the shorthand $\hat{n}(t)$ for $\widehat{Z_{n}}(t)$.

Formally, we can think of $t\in \C$ as a coordinate on the moduli space of maps from $\Sigma\cong\CP^{1}$ into $\PT$ with two fixed points (the nodes at $Z_{n}$ and $Z_{n-1}$).  We will be interested in the variation of our correlator with respect to \emph{anti-holomorphic} dependence on this coordinate; this requires a $\dbar$-operator on the moduli space $\overline{M}_{0,2}(\CP^{3|4},1)$.\footnote{These moduli spaces are, strictly speaking, algebraic stacks.  However, for the case of a genus zero Riemann surface and target space $\PT$, they are unobstructed and have a versal family which can be treated as an algebraic space.}  Formally, this can be constructed by considering the diagram:
\begin{equation*}
\xymatrix{
\overline{M}_{0,3}(\CP^{3|4},1)  \ar[d]^{\pi} \ar[rr]^{\Phi} & & \PT \\
\overline{M}_{0,2}(\CP^{3|4},1) & & }
\end{equation*}
where $\pi$ is the forgetful functor which throws away an extra marked point, and $\Phi$ is the `universal instanton.'  Since the universal curve is just $\overline{M}_{0,3}(\CP^{3|4},1)\cong \overline{M}_{0,2}(\CP^{3|4},1)\times\Sigma$, this map simply takes $f\in \overline{M}_{0,2}(\CP^{3|4},1)$ and $z\in\Sigma$ to $f(z)\in\PT$.  Hence, we can take the complex structure on $\PT$ given by $\dbar$, and define $\bar{\delta}$ on $\overline{M}_{0,2}(\CP^{3|4},1)$ both formally and heuristically:
\begin{equation*}
\bar{\delta}=\pi_{*}\Phi^{*}\dbar, \qquad \bar{\delta}=\d\bar{t}\frac{\partial}{\partial \bar{t}}.
\end{equation*}

In \cite{Bullimore:2011ni}, the twistor action and Wilson loop were used to study $\bar{\delta}\la W[C(t)]\ra$; we will use the same methodology to study the correlator between a Wilson loop and single local operator.  In direct analogy with the smooth deformation of a real Wilson loop in a real 3-manifold, Bullimore and Skinner found that the infinitesimal variation of $W[C]$ with respect to $\bar{t}$ was given by:
\be{WLvar}
\bar{\delta}\;W[C]=-\int_{C} \omega(Z)\wedge\d \bar{Z}^{\bar{\alpha}}\wedge\bar{\delta}\bar{Z}^{\bar{\beta}}\; \tr\left( F^{(0,2)}_{\bar{\alpha}\bar{\beta}}\mathrm{Hol}_{Z}[C]\right),
\ee 
where $\omega(Z)$ is a meromorphic 1-form on $C$ with simple poles at each node $Z=Z_{i}$, and $F^{(0,2)}=\dbar\cA+\cA\wedge\cA$ is the anti-holomorphic curvature of the gauge connection on twistor space.  By inserting this into the path integral for $\bar{\delta}\la W[C(t)]\ra$ with respect to the twistor action \eqref{TA}, a holomorphic analogue of the loop equations \cite{Makeenko:1979pb} was found which lead to the all-loop BCFW recursion relations of \cite{ArkaniHamed:2010kv}.

In our case, we want to consider $\bar{\delta}\la W[C(t)]\cO(y)\ra$ for any $\U(N)$ gauge group.  Since the BCFW-like deformation \eqref{BCFW1} only acts on the Wilson loop, we can use \eqref{WLvar} to consider:
\be{BCFW2}
\bar{\delta}\la W[C(t)]\cO(y)\ra =-\frac{1}{N}\int [\mathcal{D}\cA]\left[ \int_{C(t)} \omega(Z)\wedge\d \bar{Z}^{\bar{\alpha}}\wedge\bar{\delta}\bar{Z}^{\bar{\beta}} \tr\left( F^{(0,2)}_{\bar{\alpha}\bar{\beta}}\mathrm{Hol}_{Z}[C]\right) \cO(y)\right] e^{-S[\cA]},
\ee
where $\cO(y)$ is our 1/2-BPS operator \eqref{nabelian}, $S[\cA]$ is the twistor action for $\cN=4$ SYM, and we have included a normalization factor of $1/N$.  As noted earlier, the twistor action can be decomposed into a holomorphic Chern-Simons portion accounting for the SD sector of the theory (or tree-level for the Wilson loop) and a non-local contribution encoding the ASD interactions (or loop-level for the Wilson loop):
\begin{equation*}
S[\cA]=S_{1}[\cA]+\lambda S_{2}[\cA],
\end{equation*}
with
\begin{eqnarray}
S_{1}[\cA] & = & \frac{i}{2\pi}\int_{\PT}\D^{3|4}Z\wedge\tr\left(\cA\wedge\dbar\cA+\frac{2}{3}\cA\wedge\cA\wedge\cA\right) \\
S_{2}[\cA] & = & \int_{\Gamma}\d^{4|8}X\log\det\left(\dbar+\cA\right)|_{X},
\end{eqnarray}
where $\Gamma$ is a real contour over the space of all lines $X\subset\PT$ corresponding to an integral over a real slice of space-time.

\subsection{Holomorphic linking contribution}

We begin by considering the classical piece of \eqref{BCFW2} corresponding to $S_{1}[\cA]$.  In this case, we note that
\begin{equation*}
\frac{\delta S_{1}[\cA]}{\delta \cA(Z)}= N\;F^{(0,2)}(Z),
\end{equation*}
so that
\begin{equation*}
\bar{\delta}\la W[C(t)]\cO(y)\ra^{\mathrm{tree}}=\frac{1}{N^2}\int [\mathcal{D}\cA]\left[ \int_{C(t)}\omega(Z)\wedge\tr\left(\mathrm{Hol}_{Z}[C(t)]\frac{\delta}{\delta\cA(Z)}e^{-S_{1}[\cA]}\right)\cO(y)\right].
\end{equation*}
Since $S_{1}[\cA]$ corresponds to the SD sector of the theory, the path integral can be thought of as defined over the moduli space of Yang-Mills instantons on $\M$ by the Ward Correspondence \cite{Manin:1997}.  Such moduli spaces are well defined and have compactifications (c.f., \cite{Uhlenbeck:1982zm, Tian:2000fu}), so we assume that integration by parts is well-defined in the context of the path integral.  This transfers the variational derivative onto the holonomy, which gives \cite{Bullimore:2011ni}:
\begin{multline*}
\tr\left(\frac{\delta}{\delta\cA(Z)}\mathrm{Hol}_{Z}[C(t)]\right)= \\
\sum_{j=1}^{n}\int_{C_{j}(t)}\omega_{j-1,j}(Z')\wedge\bar{\delta}^{3|4}(Z,Z')\tr\left[U(Z,Z_{n})\cdots U(Z_{j},Z')\right]\tr\left[U(Z',Z_{j-1})\cdots U(Z_{1},Z)\right],
\end{multline*}
where $C_{j}(t)=(j-1,j)$ is the $j^{\mathrm{th}}$ component of the nodal curve $C(t)$; $\omega_{j-1,j}(Z)$ is the meromorphic 1-form on $C_{j}(t)$ with poles at $Z_{j-1}$, $Z_{j}$; and $\bar{\delta}^{3|4}$ is the $(0,3)$-current
\begin{equation*}
\bar{\delta}^{3|4}(Z,Z')=\int_{\C}\frac{\d s}{s}\bar{\delta}^{4|4}(Z+ sZ'),
\end{equation*}
which enforces the projective coincidence of its arguments.  

\begin{figure}
\centering
\includegraphics[width=3.5 in, height=1.5 in]{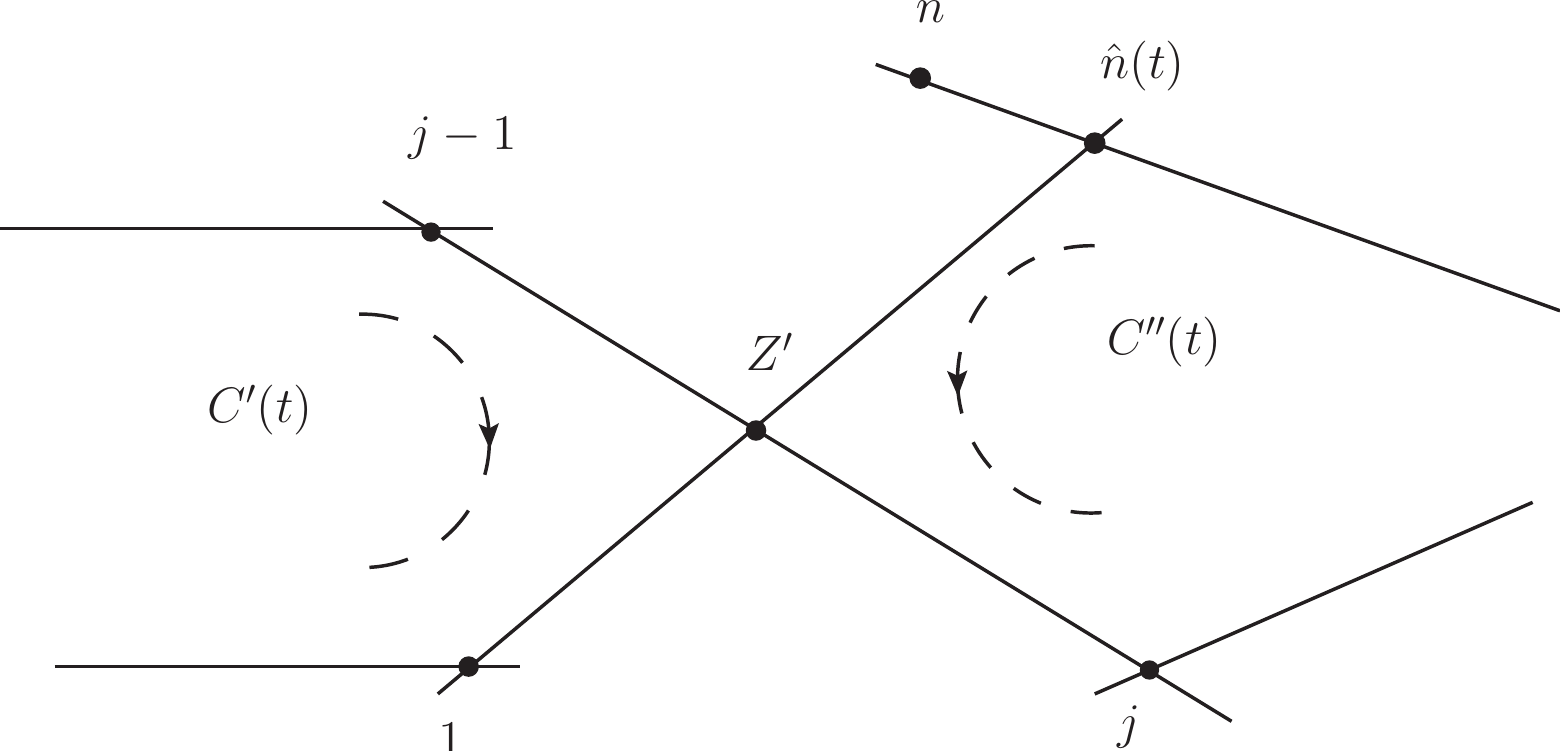}\caption{\textit{A holomorphic linking contribution when the curve $C(t)$ intersects itself.}}\label{BCF2}
\end{figure}
Indeed, $\bar{\delta}^{3|4}(Z,Z')$ indicates that $\bar{\delta}\la W[C(t)]\cO(y)\ra^{\mathrm{tree}}$ is supported only at those values of $t\in\C$ where the deformed Wilson loop intersects itself; the trace structure indicates that this results in the factorization of the original operator into two Wilson loops around the nodal curves $C'(t)$ and $C''(t)$ obtained by ungluing $C(t)$ at the new intersection point $Z=Z'$.  The geometry of this configuration is illustrated in figure \ref{BCF2}.

Thus, we have:
\be{BCFWt}
\bar{\delta}\la W[C(t)]\cO(y)\ra^{\mathrm{tree}}=- \int\limits_{C(t)\times C(t)} \omega(Z)\wedge \omega(Z')\wedge\bar{\delta}^{3|4}(Z,Z') \left\la W[C'(t)]\;W[C''(t)]\;\cO(y)\right\ra,
\ee
where we have absorbed a normalization factor of $1/N$ into each Wilson loop.  This is the analogue of the holomorphic linking term of the holomorphic loop equations derived in \cite{Bullimore:2011ni}, but now in the presence of an additional local operator in general position.  Note that in the planar limit of the gauge theory, the correlator can be re-written as
\begin{equation*}
\left\la W[C'(t)]\;W[C''(t)]\;\cO(y)\right\ra = \la W[C'(t)]\cO(y)\ra\; \la W[C''(t)]\ra + \la W[C'(t)]\ra\; \la W[C''(t)]\cO(y)\ra.
\end{equation*}

\subsection{Contributions from MHV vertices and local operator}

We must still account for the portion of $\bar{\delta}\la W[C(t)]\cO(y)\ra$ corresponding to $S_{2}[\cA]$ and the local operator $\cO(y)$ itself.  Essentially, such contributions arise because the deformed Wilson loop intersects a line $X$ corresponding to a MHV vertex from the log-det of the twistor action, or the line $Y$ of the local operator insertion.  Although the genericity assumption guarantees that no such intersections occur for $t=0$ (and furthermore that $Y$ never intersects an MHV vertex), as $t$ varies it sweeps out a plane $(n-1, n, 1)$ which any lines $X$, $Y$ in general position must intersect.  Once again, we take our cue from \cite{Bullimore:2011ni} to study such contributions to the variation of our correlator.

From the path integral, we can see that contributions from the MHV vertices must arise as:
\be{PA1}
-\frac{\lambda}{N} \int[\mathcal{D}\cA]\int_{\Gamma} \d^{4|8}X\;\left[\int_{C(t)}\omega(Z)\wedge\tr\left( \delta\log\det(\dbar+\cA)|_{X}\mathrm{Hol}_{Z}[C(t)]\right)\cO(y)\right]e^{-S[\cA]},
\ee 
with the factors of the 't Hooft coupling $\lambda$ coming from $S_{2}[\cA]$ and $1/N$ for normalization.  The variation of the log-det can be found by standard methods (c.f., \cite{Mason:2001vj}); if we assume that this line $X$ is given by the span of $Z_{A}$ and $Z_{B}$, then
\begin{equation*}
\delta\log\det(\dbar+\cA)|_{X}=\int\limits_{X\times S^{1}\times S^{1}} \omega_{A,B}(Z')\wedge \frac{\la A\d A\ra \wedge\la B \d B\ra}{\la AB\ra^{2}}\tr\left(U(Z_{B},Z')\delta\cA(Z')\right),
\end{equation*}
where $\omega_{A,B}(Z')$ is the meromorphic differential on $X$ with poles at $Z_{A}$ and $Z_{B}$, and $\lambda_{A},\lambda_{B}$ are the homogeneous coordinates of these points on $X$.  The integral over $S^{1}\times S^{1}$ is a contour integral surrounding the poles at $Z_{A}=Z_{B}=Z'$.

\begin{figure}
\centering
\includegraphics[width=4 in, height=1.5 in]{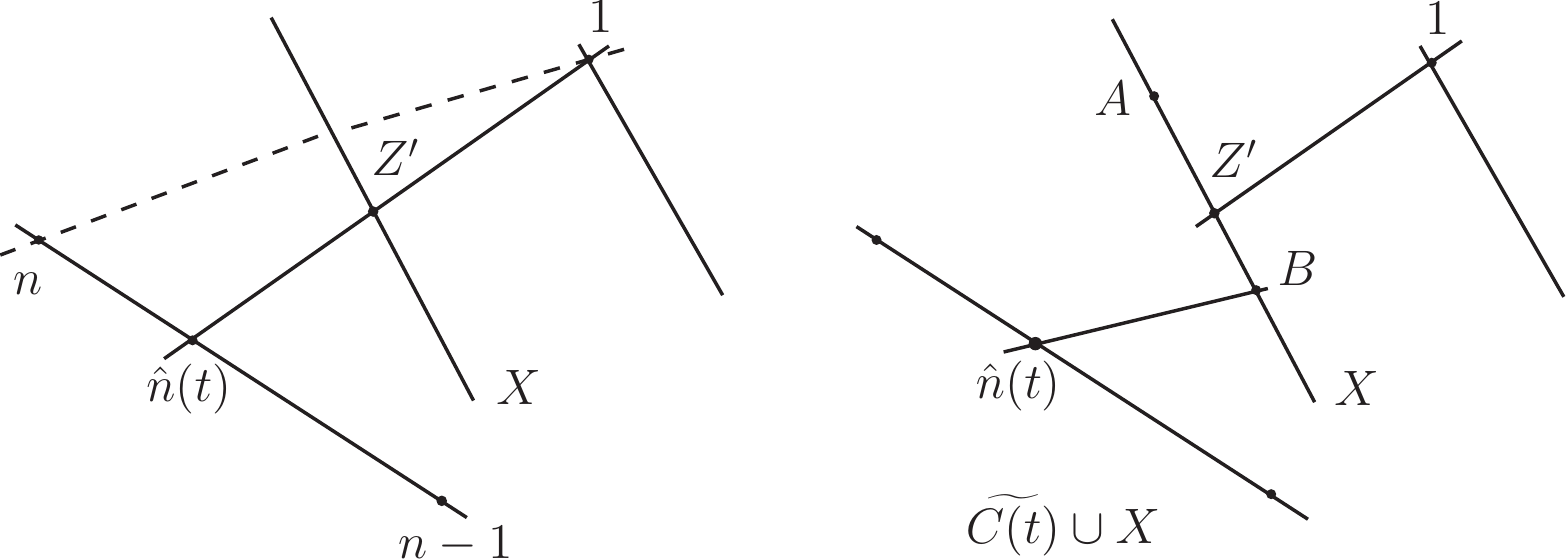}\caption{\textit{An intersection of the curve $C(t)$ with a MHV vertex $X$ (left) can be expressed as a forward limit of a new curve $\widetilde{C(t)}\cup X$ (right).}}\label{BCF3}
\end{figure}

Furthermore, the integral over the positions of $Z_{A}$ and $Z_{B}$ on $X$ can be combined with the measure $\d^{4|8}X$ to give a measure on the space of lines:
\begin{equation*}
\d^{4|8}X\wedge\frac{\la A\d A\ra \wedge\la B \d B\ra}{\la AB\ra^{2}}=\D^{3|4}Z_{A}\wedge\D^{3|4}Z_{B}.
\end{equation*}
Hence, the integrand of our path integral expression \eqref{PA1} may be written:
\begin{multline}\label{PA2}
-\frac{\lambda}{N}\oint\limits_{\Gamma\times S^{1}\times S^{1}}\D^{3|4}Z_{A}\wedge\D^{3|4}Z_{B}\:\int\limits_{C(t)\times X}\omega(Z)\wedge\omega_{A,B}(Z')\wedge\bar{\delta}^{3|4}(Z,Z')  \\
\times \tr\left(U(Z_{B},Z')\mathrm{Hol}_{Z}[C(t)]\right)\cO(y),
\end{multline}
with the $\bar{\delta}^{3|4}(Z,Z')$ ensuring that this is supported only when $C(t)$ intersects $X$ at $Z=Z'$.  As shown in \cite{Bullimore:2011ni}, this configuration can naturally be interpreted as a forward limit where the MHV vertex at $x\in\M$ becomes null separated from the point corresponding to the deformed line $(\hat{n}(t),1)$ in twistor space.  This can be operationalized by replacing $C(t)$ with a new curve $\widetilde{C(t)}$ which has an additional component such that:
\begin{equation*}
\widetilde{C(t)}\cap X=\{Z',Z_{B}\}, \qquad \lim_{Z_{B}\rightarrow Z'}\widetilde{C(t)}\rightarrow C(t),
\end{equation*}
\begin{equation*}
\widetilde{C(t)}\cup X=(1,2)\cup\cdots\cup (n-1,\hat{n}(t))\cup (Z',B)\cup (B,1).
\end{equation*}
This forward limit curve is pictured in figure \ref{BCF3}.

On the support of all the contours and delta-functions appearing in \eqref{PA2}, we can make the replacement:
\be{BCFW3}
\frac{1}{N}\tr\left(U(Z_{B},Z')\mathrm{Hol}_{Z}[C(t)]\right)=W[\widetilde{C(t)}\cup X].
\ee
This means that we get a contribution to $\bar{\delta}\la W[C(t)]\cO(y)\ra$ from the MHV vertices of the form:
\be{BCFWmhv}
-\lambda \oint\limits_{\Gamma\times S^{1}\times S^{1}}\D^{3|4}Z_{A}\wedge \D^{3|4}Z_{B} \left[ \int\limits_{C(t)\times X} \omega(Z)\wedge\omega_{A,B}(Z')\wedge\bar{\delta}^{3|4}(Z,Z')\left\la W[\widetilde{C(t)}\cup X]\;\cO(y)\right\ra \right].
\ee
Once again, this is the natural analogy of the second term in the holomorphic loop equations of \cite{Bullimore:2011ni}.

Finally, we must account for when $C(t)$ intersects $Y$, the line marking the insertion of the operator $\cO(y)$.  Clearly, the geometry of such a configuration is similar to that of the MHV vertex, although the presence of $\cO(y)$ entails a more complicated R-symmetry structure associated with the 1/2-BPS operator.  However, it can be seen that this operator intersection is covered by considering the previous case with a slight modification of the contour $\Gamma$.

Suppose that $Y$ is given by the span of $Z_{C}$ and $Z_{D}$, and consider $\delta\log\det(\dbar+\cA)|_{Y}$ as before, but without integrating fully over the position of $y\in\M$.  Instead, consider a purely fermionic integral of the form:
\begin{multline*}
- \oint\limits_{\tilde{\Gamma}\times S^{1}\times S^{1}}\d^{0|4}\theta^{abab} \wedge\frac{\la C\d C\ra\wedge\la D\d D\ra}{\la CD\ra^{2}}\\
\times\left[\int\limits_{C(t)\times Y}\omega(Z)\wedge\omega_{C,D}(Z')\wedge\bar{\delta}^{3|4}(Z,Z')\;\left\la W[\widehat{C(t)}\cup Y]\right\ra\right],
\end{multline*}
where (without loss of generality) $\d^{0|4}\theta^{abab}=\d\theta^{aA}\d\theta^{b}_{A}\d\theta^{aB}\d\theta^{b}_{B}$, $\tilde{\Gamma}$ is the corresponding fermionic contour (which, as usual, can be evaluated algebraically), and $\widehat{C(t)}$ is the forward limit curve associated with $Y$.  The R-symmetry of this measure extracts fermionic derivatives of the field $\cA$ from the holomorphic frames in $W[\widehat{C(t)}\cup Y]$, but supersymmetry dictates that they must be inserted at two different places on the line $Y$.  The remainder of the contour $\tilde{\Gamma}$ integrates these insertion points over $Y$.

Explicitly, in $W[\widehat{C(t)}\cup Y]$ we can use the properties of the holomorphic frame to write:
\begin{equation*}
U(Z_{D},Z')=U(Z_{D},Z_{C}) U(Z_{C},Z'),
\end{equation*}
and on the support of the $S^{1}\times S^{1}$ contour and $\bar{\delta}^{3|4}(Z,Z')$, we can take $Z'=Z_{D}$.  The integral over $\d^{0|4}\theta^{abab}$ then pulls two derivatives from each of these frames on $Y$, and everything remaining must be integrated over the line to give:
\begin{equation*}
\int_{Y\times Y} \la C\d C\ra\wedge\la D\d D\ra U(Z_{D},Z_{C})\frac{\partial^{2}\cA(Z_{C})}{\partial\chi^{a}\partial\chi^{b}} U(Z_{C},Z_{D})\frac{\partial^{2}\cA(Z_{D})}{\partial\chi^{a}\partial\chi^{b}}.
\end{equation*}
Of course, this is our 1/2-BPS operator $\cO(y)=\cO_{abab}(y)$ from \eqref{nabelian}, and is inserted in the color trace running over the remaining holomorphic frames of the Wilson loop at the point $Z'$.  But this is precisely what we expect for the configuration where the deformed Wilson loop $C(t)$ intersects $Y$ at the point $Z=Z'$!  In other words, the 1/2-BPS operator is also captured by the variation of the log-det term in the twistor action, but integrated over a partial fermionic contour corresponding to its R-symmetry structure.\footnote{The $\log\det(\dbar+\cA)|_{X}$ is not locally gauge invariant; it must be integrated over some contour in order to kill anomaly terms \cite{Boels:2006ir}.  The algebraic integral over $\tilde{\Gamma}$ gives the log-det precisely the gauge invariance of the 1/2-BPS operators, as desired.}

Thus we obtain a third contribution to $\bar{\delta}\la W[C(t)]\cO(y)\ra$ of the form
\be{BCFWop}
-\oint\limits_{\tilde{\Gamma}\times S^{1}\times S^{1}}\d \mu^{abab} \left[\int\limits_{C(t)\times Y}\omega(Z)\wedge\omega_{C,D}(Z')\wedge\bar{\delta}^{3|4}(Z,Z') \left\la W[\widehat{C(t)}\cup Y]\right\ra \right],
\ee
where 
\begin{equation}\label{fmeasure}
\d\mu^{abab}=\d^{0|4}\theta^{abab} \wedge\frac{\la C\d C\ra\wedge\la D\d D\ra}{\la CD\ra^{2}}.
\end{equation}

\subsection{The recursion relation}

We can now combine \eqref{BCFWt}, \eqref{BCFWmhv} and \eqref{BCFWop} to derive a BCFW-like all-loop recursion relation for the correlator $\la W[C]\cO(y)\ra$.  For this, we integrate over the moduli space coordinate $t$:
\be{recur1}
-\int_{\C}\frac{\d t}{t}\wedge\bar{\delta}\la W[C(t)]\cO(y)\ra =\int_{\C} \frac{\d t}{t}\wedge \left( \Lambda^{\mathrm{tree}}+\Lambda^{\mathrm{MHV}}+\Lambda^{\mathrm{Op}}\right),
\ee
where the $\Lambda$s are given by the deformation contributions we just calculated.  Integration by parts immediately gives
\begin{equation*}
-\int_{\C}\frac{\d t}{t}\wedge\bar{\delta}\la W[C(t)]\cO(y)\ra = \left\la W[1,2,\ldots , n]\cO(y)\right\ra-\left\la W[1,2,\ldots ,n-1]\cO(y)\right\ra,
\end{equation*}
which is just the difference in the correlators at $t=0$ and $t=\infty$.  

Now, the first contribution on the right-hand side of \eqref{recur1} is
\begin{multline*}
\int_{\C}\frac{\d t}{t}\wedge \Lambda^{\mathrm{tree}}=  \int\limits_{\C\times C(t)\times C(t)} \frac{\d t}{t}\wedge\omega(Z)\wedge\omega(Z')\wedge\bar{\delta}^{3|4}(Z,Z') \\
\times \left\la W[C'(t)]W[C''(t)]\cO(y)\right\ra.
\end{multline*}
As discussed earlier, the $\bar{\delta}^{3|4}(Z,Z')$ ensures that this is supported only when the curve $C(t)$ intersects itself.  For every $j=3,\ldots n-1$ there will be some value of $t=t_{j}$ for which the line $(\hat{n}(t_{j}),1)$ intersects $(j-1,j)$.  If we label those intersection points as $I_{j}$, then clearly we have \cite{Bullimore:2011ni}
\begin{eqnarray*}
C'(t_{j})=(1,2)\cup(2,3)\cup\cdots\cup (j-1, I_{j}) \\
C''(t_{j})=(I_{j},j)\cup(j,j+1)\cup\cdots\cup(\hat{n}(t_{j}),1).
\end{eqnarray*}
For each such contribution at $t_{j}$ we can parametrize the positions of $Z$ and $Z'$ as
\begin{equation*}
Z=\widehat{Z_{n}}(t)+sZ_{1}=Z_{n}+tZ_{n-1}+sZ_{1}, \qquad Z'=Z_{j-1}+r Z_{j},
\end{equation*}
so the meromorphic differentials become
\begin{equation*}
\omega(Z)=\frac{\d s}{s}, \qquad \omega(Z')=\frac{\d r}{r}.
\end{equation*}

Thus, we have
\begin{multline}
\int_{\C}\frac{\d t}{t}\wedge \Lambda^{\mathrm{tree}}= \sum_{j=3}^{n-1}\int_{\C^{3}}\frac{\d t}{t}\frac{\d s}{s}\frac{\d r}{r}\wedge\bar{\delta}^{3|4}(Z_{n}+tZ_{n-1}+sZ_{1},Z_{j-1}+r Z_{j}) \\
\times \left\la W[1,\ldots, j-1, I_{j}]W[I_{j},j,\ldots, n-1,\hat{n}(t_{j})]\cO(y)\right\ra \\
= \sum_{j=3}^{n-1}[n-1,n,1,j-1,j]\left\la W[1,\ldots, j-1, I_{j}]W[I_{j},j,\ldots, n-1,\hat{n}_{j}]\cO(y)\right\ra,
\end{multline}
where $[A,B,C,D,E]$ is the standard dual superconformal invariant on twistor space (the `R-invariant') \cite{Mason:2009qx}, and we have abbreviated $\hat{n}(t_{j})=\hat{n}_{j}$.  We can perform similar parametrizations for the remaining terms $\Lambda^{\mathrm{MHV}}$ and $\Lambda^{\mathrm{Op}}$.  This means we can now state our full recursion relation for the correlator:

\begin{propn}
Let $W^{n}[C]=W[1,\ldots,n]$ be the Wilson loop in the fundamental representation around the n-cusp null polygon $C$, $\cO(y)$ be a local operator in general position, and $Y=\mathrm{span}\{Z_{C},Z_{D}\}$ be the $\CP^{1}\subset\PT$ corresponding to this position.  Then
\begin{multline}\label{recur2}
\left\la W[1,\ldots, n]\cO(y)\right\ra = \left\la W[1,\ldots, n-1]\cO(y)\right\ra \\
+ \sum_{j=3}^{n-1}[n-1,n,1,j-1,j]\left\la W[1,\ldots, j-1, I_{j}]W[I_{j},j,\ldots, n-1,\hat{n}_{j}]\cO(y)\right\ra \\
+\lambda \oint\limits_{\Gamma\times S^{1}\times S^{1}}\D^{3|4}Z_{A}\wedge\D^{3|4}Z_{B} [n-1,n,1,A,B]\left\la W[1,\ldots, n-1,\hat{n}_{AB},Z',B]\cO(y)\right\ra \\
+ \oint\limits_{\tilde{\Gamma}\times S^{1}\times S^{1}} \d\mu^{abab}[n-1,n,1,C,D]\left\la W[1,\ldots, n-1, \hat{n}_{CD},Z',D]\right\ra,
\end{multline}
where the measure $\d\mu^{abab}$ is given by \eqref{fmeasure}; the contours $\Gamma$ and $\tilde{\Gamma}$ are over $(4|8)$- and $(0|4)$-dimensional real slices of the space of lines in $\PT$ respectively; and the contours $S^{1}\times S^{1}$ ensure $Z_{A,B}, Z_{C,D}\rightarrow Z'$ in their respective integrals.\footnote{As mentioned earlier, recall that in the planar limit $\la W[1,\ldots, j-1, I_{j}]W[I_{j},j,\ldots, n-1,\hat{n}_{j}]\cO(y)\ra = \la W[1,\ldots, j-1, I_{j}]\ra\; \la W[I_{j},j,\ldots, n-1,\hat{n}_{j}]\cO(y)\ra +\la W[1,\ldots, j-1, I_{j}]\cO(y)\ra\; \la W[I_{j},j,\ldots, n-1,\hat{n}_{j}]\ra$, so this indeed constitutes a recursion relation.}
\end{propn}


\section{Generalizations: Additional Operators and Null Limits}
\label{strong}

\subsection{Proofs via twistor theory}

It is natural to consider generalizations of \eqref{locW}, particularly those which include an arbitrary number of operators in general position; such extensions were proposed in \cite{Alday:2011ga} and studied at weak coupling in \cite{Engelund:2011fg}.  For instance, if we consider the following limit
\begin{equation*}
\lim_{x^{2}_{i,i+1}\rightarrow 0}\frac{\la \cO(x_{1})\cdots\cO(x_{n})\cO(y_{1})\cdots\cO(y_{k})\ra}{\la \cO(x_1)\cdots\cO(x_n)\ra},
\end{equation*}
where the $k$ operators $\cO(y_{1}),\ldots,\cO(y_k)$ remain in general position relative to the $x_{i}$ and each other, then it isn't hard to see using the methods which led to \eqref{nplim1} that this is equal to
\begin{equation*}
\frac{1}{\la W^{n}_{\mathrm{adj}}[C]\ra} \sum_{j=0}^{k-2} \sum_{\{i_{1},\ldots, i_{j}\}\subset \{1,\ldots, k\}} \la W^{n}_{\mathrm{adj}}[C]\cO(y_{i_{1}})\cdots\cO(y_{i_{j}})\ra\: \la\cO(y_{i_{j+1}})\cdots\cO(y_{i_{k}})\ra,
\end{equation*} 
where the range of the sum is dictated by normal ordering.  Taking the planar limit splits the first factor into two correlators with fundamental Wilson loops as before, and introduces another sum over partitions of the remaining operators.

An easy extension would be to consider the limit where the $k$ remaining operators become pairwise null separated themselves (separately of the first $n$); this introduces new divergences which need to be balanced by an additional denominator factor.  The natural choice is then to study the limit
\be{locW2}
\lim_{x^{2}_{i,i+1},y^{2}_{j,j+1}\rightarrow 0} \frac{\la \cO(x_{1})\cdots\cO(x_{n})\cO(y_{1})\cdots\cO(y_{k})\ra}{\la \cO(x_{1})\cdots\cO(x_{n})\ra \la\cO(y_{1})\cdots\cO(y_{k})\ra}.
\ee
Once again, we can apply the Wilson loop / correlations functions duality of \eqref{corrW} to re-write this in a more palatable form as:
\begin{equation*}
\lim_{x^{2}_{i,i+1},y^{2}_{j,j+1}\rightarrow 0} \frac{\la \cO(x_{1})\cdots\cO(x_{n})\cO(y_{1})\cdots\cO(y_{k})\ra}{\la \cO(x_{1})\cdots\cO(x_{n})\ra^{\mathrm{tree}} \la\cO(y_{1})\cdots\cO(y_{k})\ra^{\mathrm{tree}}}\frac{1}{\la W^{n}_{\mathrm{adj}}[C]\ra \la W^{k}_{\mathrm{adj}}[D]\ra}, 
\end{equation*}
where $D$ is the null polygon with $k$ cusps resulting from the additional null limit.  Once again, the tree-level correlators in the denominator go as:
\be{diverge}
\la \cO(x_{1})\cdots\cO(x_{n})\ra^{\mathrm{tree}} \la \cO(y_{1})\cdots\cO(y_{k})\ra^{\mathrm{tree}} \sim \frac{1}{x_{12}^{2}x_{23}^{2}\cdots x_{n1}^{2}} \times \frac{1}{y_{12}^{2}y_{23}^{2}\cdots y_{k1}^{2}},
\ee
so we must isolate those portions from the numerator which have a similar divergence.

To do this, we can break the numerator into a sum of connected and disconnected components of the correlation function:
\begin{multline*}
\frac{1}{\la W^{n}_{\mathrm{adj}}[C]\ra \la W^{k}_{\mathrm{adj}}[D]\ra} \lim_{x^{2}_{i,i+1},y^{2}_{j,j+1}\rightarrow 0}\left( \frac{\la \cO(x_{1})\cdots\cO(x_{n})\cO(y_{1})\cdots\cO(y_{k})\ra^{\mathrm{conn}}}{\la \cO(x_{1})\cdots\cO(x_{n})\ra^{\mathrm{tree}} \la\cO(y_{1})\cdots\cO(y_{k})\ra^{\mathrm{tree}}} \right. \\
+ \left. \frac{\la \cO(x_{1})\cdots\cO(x_{n})\ra \la\cO(y_{1})\cdots\cO(y_{k})\ra}{\la \cO(x_{1})\cdots\cO(x_{n})\ra^{\mathrm{tree}} \la\cO(y_{1})\cdots\cO(y_{k})\ra^{\mathrm{tree}}} + \frac{\{\mbox{all other disconnected}\}}{\la \cO(x_{1})\cdots\cO(x_{n})\ra^{\mathrm{tree}} \la\cO(y_{1})\cdots\cO(y_{k})\ra^{\mathrm{tree}}} \right)
\end{multline*}
We can now analyse each term by performing all contractions in twistor space and looking at their degree of divergence.  Because none of the $X_{i}$ and $Y_{j}$ ever intersect in twistor space (we assume that the two sets of operators become pairwise null separated independently), our computations in appendix \ref{calcs} indicate that the only contractions which produce the correct degree of divergence in the first term are those between $\frac{\partial^{2}\cA}{\partial \chi^{2}}$ on adjacent $X$s and adjacent $Y$s.  Hence,
\begin{equation*}
\lim_{x^{2}_{i,i+1},y^{2}_{j,j+1}\rightarrow 0}\frac{\la \cO(x_{1})\cdots\cO(x_{n})\cO(y_{1})\cdots\cO(y_{k})\ra^{\mathrm{conn}}}{\la \cO(x_{1})\cdots\cO(x_{n})\ra^{\mathrm{tree}} \la\cO(y_{1})\cdots\cO(y_{k})\ra^{\mathrm{tree}}}=\la W^{n}_{\mathrm{adj}}[C] W^{k}_{\mathrm{adj}}[D]\ra^{\mathrm{conn}}.
\end{equation*}

The second term is easily evaluated by again applying \eqref{corrW}:
\begin{equation*}
\lim_{x^{2}_{i,i+1},y^{2}_{j,j+1}\rightarrow 0}\frac{\la \cO(x_{1})\cdots\cO(x_{n})\ra \la\cO(y_{1})\cdots\cO(y_{k})\ra}{\la \cO(x_{1})\cdots\cO(x_{n})\ra^{\mathrm{tree}} \la\cO(y_{1})\cdots\cO(y_{k})\ra^{\mathrm{tree}}}=\la W^{n}_{\mathrm{adj}}[C]\ra \la W^{k}_{\mathrm{adj}}[D]\ra .
\end{equation*}
The remaining terms (composed of all other disconnected components from the correlation function) involve all the usual contractions which give a vanishing contribution (e.g., contractions between non-adjacent lines, contractions between operators and fields on any line with a MHV vertex), but in addition contain no connected component with enough lines in twistor space to form a full Wilson loop in the null limit.  So any term in this sum of disconnected components will contain some divergences of the form $x_{i,i+1}^{-2} y_{j,j+1}^{-2}$, but never the full array appearing in \eqref{diverge}.  Thus, all remaining disconnected terms vanish in the null limit, allowing us to write:
\begin{equation*}
\lim_{x^{2}_{i,i+1},y^{2}_{j,j+1}\rightarrow 0} \frac{\la \cO(x_{1})\cdots\cO(x_{n})\cO(y_{1})\cdots\cO(y_{k})\ra}{\la \cO(x_{1})\cdots\cO(x_{n})\ra \la\cO(y_{1})\cdots\cO(y_{k})\ra} = 1+\frac{\la W^{n}_{\mathrm{adj}}[C] W^{k}_{\mathrm{adj}}[D]\ra^{\mathrm{conn}}}{\la W^{n}_{\mathrm{adj}}[C]\ra \la W^{k}_{\mathrm{adj}}[D]\ra}.
\end{equation*}

In both of the generalizations we have discussed here, the passage to the planar limit of the gauge theory happens as in proposition \ref{locP1}, allowing us to state our full generalizations as follows:
\begin{corol}\label{locP2}
Let $\{\cO(x_{i}), \cO(y_{j})\}^{i=1,\ldots,n}_{j=1,\ldots,k}$ be gauge invariant local operators in $\cN=4$ SYM, $C$ be the null polygon resulting from the limit where the $\{\cO(x_{i})\}$ become pairwise null separated (i.e., $x_{i,i+1}^{2}=0$), and $D$ be the null polygon when the $\{\cO(y_{j})\}$ become null separated ($y^{2}_{j,j+1}=0$).  Then at the level of the integrand,
\begin{multline}\label{locW2}
\lim_{x^{2}_{i,i+1}\rightarrow 0}\frac{\la \cO(x_{1})\cdots\cO(x_{n})\cO(y_{1})\cdots\cO(y_{k})\ra}{\la \cO(x_1)\cdots\cO(x_n)\ra} \\
= \frac{1}{\la W^{n}_{\mathrm{adj}}[C]\ra} \sum_{j=0}^{k-2} \sum_{\{i_{1},\ldots, i_{j}\}\subset \{1,\ldots, k\}} \la W^{n}_{\mathrm{adj}}[C]\cO(y_{i_{1}})\cdots\cO(y_{i_{j}})\ra\: \la\cO(y_{i_{j+1}})\cdots\cO(y_{i_{k}})\ra \\
\xrightarrow{\mathrm{planar}\:\mathrm{limit}} \frac{1}{\la W^{n}[C]\ra^{2}} \sum_{\cP_{k}} \la W^{n}[C]\cO(y_{i_{1}})\cdots\cO(y_{i_{j}})\ra \;\la W^{n}[C]\cO(y_{i_{j+1}})\cdots\cO(y_{i_{l}})\ra \\
\times \la \cO(y_{i_{l+1}})\cdots\cO(y_{i_{k}})\ra,
\end{multline}
where all expectation values are assumed to be generic and normal ordered, $W^{n}[C]$ is the Wilson loop in the fundamental representation, and $\sum_{\cP_{k}}$ is the sum over relevant partitions of $\{1,\ldots,k\}$.  If we allow the remaining $k$ correlation functions to also become null separated, then
\begin{multline}\label{locW3}
\lim_{x^{2}_{i,i+1},y^{2}_{j,j+1}\rightarrow 0} \frac{\la \cO(x_{1})\cdots\cO(x_{n})\cO(y_{1})\cdots\cO(y_{k})\ra}{\la \cO(x_{1})\cdots\cO(x_{n})\ra \la\cO(y_{1})\cdots\cO(y_{k})\ra} = 1+\frac{\la W^{n}_{\mathrm{adj}}[C] W^{k}_{\mathrm{adj}}[D]\ra^{\mathrm{conn}}}{\la W^{n}_{\mathrm{adj}}[C]\ra \la W^{k}_{\mathrm{adj}}[D]\ra} \\
\xrightarrow{\mathrm{planar}\:\mathrm{limit}} 1+2\frac{\la W^{n}[C] W^{k}[D]\ra^{\mathrm{conn}}}{\la W^{n}[C]\ra \la W^{k}[D]\ra}.
\end{multline}
\end{corol}

\subsection{An example: Two operators in general position}

The authors of \cite{Alday:2011ga} studied the correlator
\begin{equation*}
\cC^{n}_{1}(W^{n},y)=\frac{\la W^{n}[C] \cO(y)\ra}{\la W^{n} \ra}
\end{equation*}
in both the strong and weak coupling regimes, when $\cO$ was the dilaton operator or chiral primary operator.  By considering the dependence of $\cC^{n}_{1}$ on conformal cross-ratios and using overall conformal invariance, they were able to write an ansatz for the functional form of $\cC^{n}_{1}$ which contained a single undetermined function $f$ of the conformal cross-ratios.  In particular, when $n=4$, the minimal surface solution for the 4-cusp Wilson loop in $AdS_{5}\times S^{5}$ is known \cite{Alday:2007hr} and explicit strong coupling calculations are possible.  In this setting, $f$ is a function of a single conformal cross-ratio and it can be determined precisely in the strong coupling regime using a semi-classical string theory approximation in the AdS-geometry \cite{Berenstein:1998ij, Zarembo:2002ph, Roiban:2010fe, Zarembo:2010rr, Costa:2010rz, Buchbinder:2010ek}; this method has also been used to study a similar correlator involving a circular Wilson loop (i.e., $n\rightarrow\infty$) \cite{Alday:2011pf}. 

In this section, we use corollary \ref{locP2} to perform a simple analysis along similar lines when there are two operators in general position.  We begin by considering any local operators in $\cN=4$ SYM, but later work with the dilaton operator to obtain some explicit results.  Note that throughout we work in the planar limit of the gauge theory.

\subsubsection{The structure of the correlator}

By \eqref{locW2}, we have that in the planar limit
\begin{multline*}
\lim_{x_{i,i+1}^{2}\rightarrow 0} \frac{\la\cO(x_{1})\cdots\cO(x_{n})\cO(y_{1})\cO(y_{2})\ra}{\la\cO(x_{1})\cdots\cO(x_{n})\ra}=2\frac{\la W^{n}[C]\cO(y_{1})\ra \la W^{n}[C]\cO(y_{2})\ra}{\la W^{n}[C]\ra^{2}}+\la\cO(y_1)\cO(y_2)\ra \\
 +2\frac{\la W^{n}[C]\cO(y_1)\cO(y_2)\ra}{\la W^{n}[C]\ra},
\end{multline*}
with all Wilson loops in the fundamental representation.  Assuming that the first two terms are well understood by \cite{Alday:2011ga} or standard quantum field theory respectively, the third term will serve as the correlator we wish to study:
\be{2corr}
\cC^{n}_{2}(W^{n}, y_{1},y_{2})= \frac{\la W^{n}[C]\cO(y_1)\cO(y_2)\ra}{\la W^{n}[C]\ra}.
\ee
Conformal invariance dictates that this correlator should depend only on conformal cross-ratios involving the positions of the $n$ cusps of the Wilson loop $W^{n}[C]$ and the two operator locations $y_{1}$ and $y_{2}$ \cite{Drummond:2007au}.  

Beginning with the correlation function 
\begin{equation*}
\frac{\la\cO(x_{1})\cdots\cO(x_{n})\cO(y_{1})\cO(y_{2})\ra}{\la\cO(x_{1})\cdots\cO(x_{n})\ra},
\end{equation*}
the configuration of points $\{x_{1},\ldots,x_{n},y_{1},y_{2}\}$ completely beaks the conformal group provided $n>2$ \cite{Alday:2009dv}.  The $n+2$ operator positions give $4(n+2)$ total coordinates (we work bosonically at this level), so from this we must subtract the dimension of the conformal group $\SU(2,2)$, which is $15$, leaving $4n-7$ parameters upon which the correlation function can depend.  The null limit required to arrive at $\cC^{n}_{2}$ adds a light-like constraint for every cusp of the resulting Wilson loop; this means that $\cC^{n}_{2}$ can be a function of only
\be{crossratio}
c^{n}_{2}=3n-7
\ee
conformal cross-ratios.  We now want to use conformal invariance to fix $\cC^{n}_{2}$ up to a single function of these cross-ratios.

By definition \eqref{2corr}, $\cC^{n}_{2}$ should behave like the product $\cO(y_1)\cO(y_2)$ under the conformal group.  Since dilatations and inversions (together with translations) generate special conformal transformations, it suffices to study the behavior of the correlator under these transformations:
\begin{equation*}
\mathrm{dilatations}\:\left\{
\begin{array}{c}
x_{i}\rightarrow h x_{i}, \\
y_{1,2}\rightarrow h y_{1,2}
\end{array} \right. \qquad \Rightarrow \qquad \cC^{n}_{2}\rightarrow h^{-2\Delta}\cC^{n}_{2},
\end{equation*}
\begin{equation*}
\mathrm{inversions}\:\left\{
\begin{array}{c}
x^{\mu}_{i}\rightarrow\frac{x_{i}^{\mu}}{x_{i}^2}, \\
y^{\mu}_{1,2}\rightarrow\frac{y_{1,2}^{\mu}}{y_{1,2}^2}
\end{array} \right. \qquad \Rightarrow \qquad \cC^{n}_{2}\rightarrow |y_{1}|^{2\Delta}|y_{2}|^{2\Delta}\cC^{n}_{2},
\end{equation*}
where $\Delta$ is the scaling dimension of the operator $\cO$.

The asymptotic (i.e., large $y_{1}$ or $y_{2}$) behavior of $\cC^{n}_{2}$ should be governed by the operator product expansion (OPE) of the Wilson loop, as in \cite{Alday:2011ga}.  Provided the area of the null polygon $C$ is not too large (in the sense of \cite{Shifman:1980ui}; i.e., in the Wick-rotated Euclidean sense), we have that \cite{Berenstein:1998ij}
\be{OPE}
\frac{W^{n}[C]}{\la W^{n}[C]\ra}=1+\sum_{k\in \{\mathrm{CPO}\}}c_{k}A^{2\Delta_{k}}\cO_{k}(0)+\left\{\mbox{conformal descendants}\right\},
\ee
where the sum runs over all conformal primary operators (CPOs), and $A$ is the area of the disc bounded by $C$.  Hence, we have
\begin{equation*}
\frac{\la W^{n}[C]\cO(y_1)\cO(y_2)\ra}{\la W^{n}[C]\ra}|_{|y_{1,2}|\rightarrow \infty}\sim \la\cO^{\dagger}(0)\cO(y_{1,2})\ra \sim \frac{1}{|y_{1,2}|^{2\Delta}}.
\end{equation*}
Furthermore, as long as the operators under consideration are identical, the correlator must be symmetric under exchange of $y_{1}$ and $y_{2}$.  This allows us to make an ansatz for the form of $\cC^{n}_{2}$ which is the natural generalization of the one used by Alday, Buchbinder and Tseytlin in \cite{Alday:2011ga}:
\be{a1}
\cC^{n}_{2}(W^{n},y_{1},y_{2})=\frac{\mathcal{F}(y_{1},y_{2},x_{1},\ldots, x_{n})}{\prod_{i=1}^{n}|y_{1}-x_{i}|^{\frac{2\Delta}{n}} \prod_{j=1}^{n}|y_{2}-x_{j}|^{\frac{2\Delta}{n}}}.
\ee

Of course, the conformal symmetry constraints on $\cC^{n}_{2}$ allow us to fix the transformation properties of $\mathcal{F}$ as well:
\begin{equation*}
\mathrm{dilatations}\: :\: \mathcal{F}\rightarrow h^{2\Delta}\mathcal{F}, \qquad \mathrm{inversions}\: : \: \mathcal{F}\rightarrow \left(|x_{1}|\cdots |x_{n}|\right)^{-\frac{4}{n}\Delta}\mathcal{F}.
\end{equation*}
We can use these constraints to make an additional ansatz which fixes the dependence of $\mathcal{F}$ on the $\frac{n}{2}(n-3)$ non-adjacent distances $|x_{i}-x_{j}|$ and preserves all the correct behavior for our correlator:
\be{a2}
\cC^{n}_{2}(W^{n}[C],y_{1},y_{2})=\frac{\prod_{i<j-1}^{n} |x_{i,j}|^{\frac{4\Delta}{n(n-3)}} F(\zeta_{1},\ldots, \zeta_{3n-7})}{\prod_{i=1}^{n}|y_{1}-x_{i}|^{\frac{2\Delta}{n}} \prod_{j=1}^{n}|y_{2}-x_{j}|^{\frac{2\Delta}{n}}},
\ee
where the $\{\zeta_{i}\}_{i=1,\ldots , 3n-7}$ are the conformal cross-ratios containing the remaining degrees of freedom in $\cC^{n}_{2}$.  As discussed in \cite{Alday:2011ga} for the case of a single operator, both $F$ and $\Delta$ will (generally) be functions of the 't Hooft coupling $\lambda$; however, we conjecture that (as in the case of a single operator) the general structure of \eqref{a2} should hold for all values of the coupling.

To find evidence for this claim, we would like to employ a semi-classical string theory calculation in $AdS_{5}\times S^{5}$; however, we will see that such a computation is actually governed by $\cC^{n}_{1}$.

\subsubsection{Conformal cross-ratios for the regular $n=4$ Wilson loop}

We now attempt to verify the ansatz \eqref{a2} by using a semi-classical approximation analogous to that used in \cite{Alday:2011ga} for the case of $\cC^{n}_{1}$.  Using the simple case of the regular $n=4$ cusp Wilson loop, we can confirm the functional form of the ansatz, but cannot fully determine the dependence on conformal cross-ratios.  By \eqref{crossratio}, it is clear that $\cC^{4}_{2}$ will depend on five independent conformal cross-ratios, which can be written as:
\begin{eqnarray*}
\zeta_{1}=\frac{|y_{1}-x_{2}|^{2}|y_{1}-x_{4}|^{2} |x_{13}|^{2}}{|y_{1}-x_{1}|^{2}|y_{1}-x_{3}|^{2}|x_{12}|^{2}}, \qquad \zeta_{2}=\frac{|y_{2}-x_{2}|^{2}|y_{2}-x_{4}|^{2} |x_{13}|^{2}}{|y_{2}-x_{1}|^{2}|y_{2}-x_{3}|^{2}|x_{12}|^{2}}, \\
\zeta_{3}=\frac{|y_{1}-x_{2}|^{2}|y_{2}-x_{4}|^{2} |x_{13}|^{2}}{|y_{1}-x_{1}|^{2}|y_{2}-x_{3}|^{2}|x_{12}|^{2}}, \qquad \zeta_{4}=\frac{|y_{2}-x_{2}|^{2}|y_{1}-x_{4}|^{2} |x_{13}|^{2}}{|y_{2}-x_{1}|^{2}|y_{1}-x_{3}|^{2}|x_{12}|^{2}},
\end{eqnarray*}
\begin{equation*}
\zeta_{5}=\frac{|y_{1}-x_{2}|^{2}|y_{2}-x_{4}|^{2} |x_{13}|^{2}}{|y_{2}-x_{1}|^{2}|y_{1}-x_{3}|^{2}|x_{12}|^{2}}.
\end{equation*}

Using the known solution for the 4-cusp null polygonal Wilson loop as a minimal surface on in $AdS_{5}$, we can determine these conformal cross-ratios explicitly.  Recall that in the Poincar\'{e} coordinate patch, the metric on $AdS_{5}$ takes the form:
\begin{equation*}
\d s^{2}_{AdS_{5}}=R^{2}\left(\frac{\d s^{2}_{\M}+\d z^{2}}{z^{2}}\right),
\end{equation*}
where $z$ is the radial direction away from the horizon and the Minkowskian metric portion is charted with coordinates $x^{\mu}=(x^{0},x^{1},x^{2},x^{3})$ (c.f., \cite{Aharony:1999ti}).  For the regular (i.e., equal-sided) 4-cusp Wilson loop, the classical Euclidean world-sheet solution is given by \cite{Alday:2007hr}
\begin{equation*}
z=\frac{r}{\cosh(u)\cosh(v)}, \qquad x^{\mu}=\left(r\tanh(u)\tanh(v),\; r\tanh(u),\; r\tanh(v),\; 0\right),
\end{equation*}
with $u, v \in\R$, and $r$ the scale of the Wilson loop (roughly, the radius of the disc bounded by $C$).\footnote{The analysis for the irregular 4-cusp Wilson loop follows similarly via the methods of \cite{Alday:2011ga}.}  Without loss of generality, we can set $r=1$, and the position of the cusps are given when $u,v\rightarrow\pm\infty$:
\begin{eqnarray*}
x_{1}=(1,1,1,0), & x_{2}=(-1,1,-1,0), \\
x_{3}=(1,-1,-1,0), & x_{4}=(-1,-1,1,0),   
\end{eqnarray*}
at $z=0$ (i.e., on the horizon).  Feeding these back into our expressions for the conformal cross-ratios above gives exact formulae.  With the notation
\begin{equation*}
q=1-(y_{1}^{0})^{2}+(y_{1}^{1})^{2}+(y_{1}^{2})^{2}+(y_{1}^{3})^{2}, \qquad s=1-(y_{2}^{0})^{2}+(y_{2}^{1})^{2}+(y_{2}^{2})^{2}+(y_{2}^{3})^{2},
\end{equation*}
we find 
\begin{eqnarray*}
\zeta_{1}=\frac{(\frac{q}{2}-y_{1}^{0}-y_{1}^{1}+y_{1}^{2})(\frac{q}{2}-y_{1}^{0}+y_{1}^{1}-y_{1}^{2})}{(\frac{q}{2}+y_{1}^{0}-y_{1}^{1}-y_{1}^{2})(\frac{q}{2}+y_{1}^{0}+y_{1}^{1}+y_{1}^{2})}, \qquad \zeta_{2}=\frac{(\frac{s}{2}-y_{2}^{0}-y_{2}^{1}+y_{2}^{2})(\frac{s}{2}-y_{2}^{0}+y_{2}^{1}-y_{2}^{2})}{(\frac{s}{2}+y_{2}^{0}-y_{2}^{1}-y_{2}^{2})(\frac{s}{2}+y_{2}^{0}+y_{2}^{1}+y_{2}^{2})}, \\
\zeta_{3}=\frac{(\frac{q}{2}-y_{1}^{0}-y_{1}^{1}+y_{1}^{2})(\frac{s}{2}-y_{2}^{0}+y_{2}^{1}-y_{2}^{2})}{(\frac{q}{2}+y_{1}^{0}-y_{1}^{1}-y_{1}^{2})(\frac{s}{2}+y_{s}^{0}+y_{s}^{1}+y_{s}^{2})}, \qquad \zeta_{4}=\frac{(\frac{s}{2}-y_{s}^{0}-y_{s}^{1}+y_{s}^{2})(\frac{q}{2}-y_{1}^{0}+y_{1}^{1}-y_{1}^{2})}{(\frac{s}{2}+y_{2}^{0}-y_{2}^{1}-y_{2}^{2})(\frac{q}{2}+y_{1}^{0}+y_{1}^{1}+y_{1}^{2})},
\end{eqnarray*}
\begin{equation*}
\zeta_{5}=\frac{(\frac{q}{2}-y_{1}^{0}-y_{1}^{1}+y_{1}^{2})(\frac{s}{2}-y_{2}^{0}+y_{2}^{1}-y_{2}^{2})}{(\frac{s}{2}+y_{2}^{0}-y_{2}^{1}-y_{2}^{2})(\frac{q}{2}+y_{1}^{0}+y_{1}^{1}+y_{1}^{2})}.
\end{equation*}

\subsubsection{The dilaton operator at strong coupling}

For concreteness, let us take $\cO=\cO_{\mathrm{dil}}$, the dilaton operator in $\cN=4$ SYM.  As we saw in \ref{dilmot}, this is the operator which originally motivated the conjecture \eqref{locW1} and it has been studied extensively in the context of $AdS_{5}\times S^{5}$ string theory, making it a natural candidate for explicit calculations.  Any local operator $\cO(y)$ is represented in string theory by a marginal vertex operator $V(y)$, which is the integral over the world-sheet (topologically a disc in the planar limit) of a suitable integrand \cite{Polyakov:2001af, Tseytlin:2003ac}.  In the case of the dilaton, we can write this vertex operator explicitly \cite{Roiban:2010fe}:
\begin{multline}\label{dilaton2}
V_{\mathrm{dil}}(y)=c_{\mathrm{dil}}\int_{\Sigma}\d^{2}\xi \left(\frac{z}{z^{2}+(x-y)^{2}}\right)^{\Delta(J_{S^{1}})}\left(\cos(\theta)e^{i\phi}\right)^{J_{S^1}}\mathcal{L}_{AdS_{5}\times S^{5}} \\
=\int_{\Sigma}\d^{2}\xi\; \mathcal{V}_{\mathrm{dil}}[X(\xi); y],
\end{multline}
where $J_{S^1}$ is an angular momentum along a $S^{1}\subset S^{5}$, $\Delta(J_{S^{1}})=4+J_{S^1}$ is the scaling dimension, $\mathcal{L}_{AdS_{5}\times S^{5}}$ is the $AdS_{5}\times S^{5}$ Lagrangian, and $\Sigma$ is the disc-like world-sheet with $\partial \Sigma=C$.  We abbreviate all of the fields of the $AdS_{5}\times S^{5}$ superstring on the world-sheet as $X(\xi)$ in the last line.  For simplicity, we consider the case where $J_{S^1}=0$; then the normalization coefficient is \cite{Berenstein:1998ij, Zarembo:2010rr}
\begin{equation*}
c_{\mathrm{dil}}=\frac{\sqrt{6\lambda}}{8\pi N}.
\end{equation*}
Hence, we have:
\be{dcorr1}
\cC^{4}_{2\;\mathrm{dil}}(W^{4}, y_{1}, y_{2})=\frac{1}{\la W^{4}[C]\ra}\int [\mathcal{D}X]V_{\mathrm{dil}}(y_{1})V_{\mathrm{dil}}(y_{2})e^{-\frac{\sqrt{\lambda}}{2\pi} S_{AdS_{5}\times S^{5}}[X]}.
\ee

The semi-classical approximation of this correlator is then reached by taking the strong coupling limit where $\sqrt{\lambda}\gg 1$ and assuming that the scaling dimensions and charges of $V_{\mathrm{dil}}$ are much smaller than $\sqrt{\lambda}$.  In this case, the leading contribution to the path integral \eqref{dcorr1} is the one in which the world-sheet is un-distorted by the operator insertions \cite{Roiban:2010fe}, so we have:
\begin{multline}\label{dcorr2}
\lim_{\sqrt{\lambda}\rightarrow\infty}\cC^{4}_{2\;\mathrm{dil}}(W^{4}, y_{1}, y_{2})=\frac{V_{\mathrm{dil}}(y_{1})V_{\mathrm{dil}}(y_{2})\la W^{4}[C]\ra}{\la W^{4}[C]\ra} \\
= \left(\int_{\Sigma}\d^{2}\xi\; \mathcal{V}_{\mathrm{dil}}[X(\xi); y_{1}]\right)\left(\int_{\Sigma}\d^{2}\xi\; \mathcal{V}_{\mathrm{dil}}[X(\xi); y_{2}]\right).
\end{multline}
But this puts us back into the setting of \cite{Alday:2011ga}, where the semi-classical approximation of the correlator between the 4-cusp Wilson loop and a single dilaton operator was studied.  Since that correlator was dependent upon only a single conformal cross-ratio, Alday, Buchbinder and Tseytlin were able to solve for its functional form explicitly, and we can simply apply their results twice to get the leading semi-classical contribution to $\cC^{4}_{2}$:
\be{dcorr3}
\cC^{4}_{2\;\mathrm{dil}}(W^{4}, y_{1}, y_{2})|_{\mathrm{semi-cl}}\sim \frac{\prod_{i<j-1}^{4}|x_{i,j}|^{4} f(\zeta_{1})\;f(\zeta_{2})}{\prod_{i=1}^{4}|y_{1}-x_{i}|^{2} \prod_{j=1}^{4}|y_{2}-x_{j}|^{2}},
\ee
where the function $f$ is given by \cite{Alday:2011ga}:
\begin{equation*}
f(\zeta)=\frac{c_{\mathrm{dil}}\:\zeta}{3(\zeta -1)^{3}}\left[(\zeta +1)\log\zeta -2(\zeta -1)\right].
\end{equation*}

While this confirms our ansatz \eqref{a2} at leading order in the semi-classical approximation, it is dramatically less informative than the analogous calculation for $\cC^{4}_{1\;\mathrm{dil}}$ performed by Alday, Buchbinder and Tseytlin.  In that case, the correlator depended on a single conformal cross-ratio, and the only undetermined function was $f(\zeta)$; this was completely fixed at strong coupling using the semi-classical approximation.  In the case of $\cC^{4}_{2}$, we see that the leading behavior of the correlator depends only on the cross-ratios $\zeta_{1}$, $\zeta_{2}$ indicating that at strong coupling, the correlator is only sensitive to the independent locations of the operator insertions.  Indeed, this is the behavior predicted by \cite{Alday:2011ga}.  To determine the dependence on the remaining cross-ratios ($\zeta_{3}$, $\zeta_{4}$, $\zeta_{5}$), one must compute the sub-leading terms in the strong coupling expansion.


\section{Discussion \& Conclusion}
\label{conclusion}

By considering the conjecture of Alday, Buchbinder and Tseytlin in twistor space, we have seen that a proof at the level of the integrand for any gauge invariant local operators in $\cN=4$ SYM is relatively easy.  This demonstrates that twistor theory remains a useful tool even in settings where some local operators remain in general position.  Furthermore, by studying gauge theory on twistor space (in particular via the twistor action for $\cN=4$ SYM), we can find interesting recursion relations for correlators involving null polygonal Wilson loops and these local operators.  It is worth noting that while we derived this BCFW-like recursion in \S \ref{BCFW} for the case of a single local operator, it generalizes in the obvious way to include an arbitrary number of operators in general position (as studied in \S \ref{strong}), and future research should test whether or not our recursion relation could be used to calculate new quantities (even at tree-level) explicitly.

While twistor theory gives elegant descriptions of supersymmetry and null geometry, a clear drawback to this approach is the fact that it works only at the level of the \emph{loop integrand}.  To extend such results to full loop integrals, a coherent regularization scheme is required to deal with IR divergences.  Very recently, the conjectures we considered in this paper have been confirmed at weak coupling by Engelund and Roiban \cite{Engelund:2011fg} using dimensional regularization.  Though by far the most common regulator, dimensional regularization is un-physical, breaks the dual superconformal symmetry of $\cN=4$ SYM (and hence the Yangian symmetry algebra), and furthermore cannot be implemented on twistor space.  A much more physical regulator is the massive (or Higgs) regulation scheme of \cite{Alday:2009zm}, which not only preserves dual superconformal symmetry at loop level \cite{Henn:2011xk} but also shows promise for being adapted to the twistor framework.  A natural question to ask is then: can a weak-coupling analysis similar to that of \cite{Engelund:2011fg} confirm the conjectures studied in this paper using the massive regularization scheme?  Of course, to extend this question to the twistorial setting, one must first provide a method for implementing the massive regularization mechanism for the loop integrand in twistor space.

On a different note, other open questions remain at strong coupling.  As we saw in \S \ref{strong}, the semi-classical analysis used in \cite{Alday:2011ga} fails to provide much meaningful information about correlators with Wilson loops and more than one local operator, even in the simple 4-cusp case.  To gain more insight, one must compute sub-leading terms in the strong coupling expansion, but even then it may be very difficult to determine the dependence of the correlator on conformal cross-ratios: indeed even the 4-cusp, 2-operator case we studied will depend on five such variables.  As suggested by \cite{Alday:2011ga}, a more interesting challenge may be to study the cases with more cusps, attempting to gain some insight into how to solve the minimal surface equations in $AdS_{5}\times S^{5}$ using integrability.

\acknowledgments

It is my pleasure to thank Fernando Alday, Mat Bullimore, Ron Reid-Edwards, Radu Roiban, and Dave Skinner for many useful conversations and for their comments on earlier drafts of this paper.  This work was supported by a National Science Foundation (USA) Graduate Research Fellowship and by Balliol College.

\appendix

\section{Computing Contractions in Twistor Space}
\label{calcs}

This appendix contains the explicit calculations which constitute the proof of proposition \ref{locP1} as well as corollary \ref{locP2}.  All contractions are performed using the twistor propagator in CSW gauge \eqref{tprop}
\begin{equation*}
\Delta_{*}(Z,Z')=\bar{\delta}^{2|4}(Z, *, Z')=\int_{\C^{2}}\frac{\d s}{s}\frac{\d t}{t} \bar{\delta}^{4|4}(Z +s Z_{*} +t Z'),
\end{equation*}
where we suppress gauge indices, which are irrelevant until one considers the planar limit.  Recall that in twistor space, we have $n$ projective lines $X_{i}$ which correspond to the $n$ cusps obtained in the null limit.  These lines can be parametrized by $Z(s_{i})=Z_{A_i}+s_{i}Z_{B_i}$, with $s_{i}$ acting as an inhomogeneous coordinate on $X_{i}$.  The homogeneous measure $\la\lambda_{i}\d\lambda_{i}\ra$ is then written as $\la A_{i} B_{i}\ra \d s_{i}$.  Without loss of generality, we can choose $Z_{A_i}$ and $Z_{B_i}$ to be the intersection points that $X_{i}$ develops with $X_{i-1}$ and $X_{i+1}$ respectively in the null limit (i.e., as $x_{i,i+1}^{2}\rightarrow 0$, $Z_{B_{i}}\rightarrow Z_{A_{i+1}}$).  Similarly, we parametrize the line $Y$ (corresponding to the operator in general position) by $Z(\sigma)= Z_{C}+\sigma Z_{D}$, and a line $X$ corresponding to an arbitrary MHV vertex from the twistor action as $Z(t)=Z_{A}+t Z_{B}$.  Without loss of generality, we assume that the fixed CSW reference twistor $Z_{*}$ has no fermionic part (i.e., $\chi_{*}=0$).

It is useful to note that for generic $Z(s)=Z_{A}+s Z_{B}$ and $Z(t)=Z_{C}+t Z_{D}$, the basic properties of distributional forms (c.f., \cite{Adamo:2011cb, Adamo:2011pv}) indicate that
\begin{equation*}
\int_{\C^{2}}\frac{\d s}{s}\frac{\d t}{t}\bar{\delta}^{2|4}(Z(s), *, Z(t))= [A,B,*,C,D],
\end{equation*}
where the right-hand side is the standard dual superconformal invariant, or R-invariant, given by \cite{Mason:2009qx}:
\be{r-invariant}
[1,2,3,4,5]=\int_{\C^{4}}\frac{\d^{4}s}{s_{1}s_{2}s_{3}s_{4}}\bar{\delta}^{4|4}\left(Z_{1}+\sum_{i=1}^{4}s_{i}Z_{i}\right)=\frac{\delta^{0|4}((1234)\chi_{5}+\mathrm{cyclic})}{(1234)(2345)(3451)(4512)(5123)},
\ee
with $(1234)=\epsilon_{\alpha\beta\gamma\delta}Z^{\alpha}_{1}Z^{\beta}_{2}Z^{\gamma}_{3}Z^{\delta}_{4}$.  

We now begin computing each class of contractions needed for the proof of proposition \ref{locP1}

\subsection*{\textit{Contractions involving an MHV vertex}}

Consider an arbitrary MHV vertex supported on $X$ and the operators and frame supported on any of the $X_{i}$.  First, take the contraction between a field $\cA$ inside one of the holomorphic frames $U_{X_i}$ and a field $\cA$ on $X$; this is given by:
\be{C1}
\left\la \overbrace{\cA|_{X_{i}}\cA|_{X}} \right\ra=\int_{\C^{2}} \frac{\d s_{i}}{s_{i}}\frac{\d t}{t} \Delta_{*}(Z(s_{i}), Z(t))=[A_{i},B_{i},*,A,B].
\ee
The genericity assumption means that (even in the null limit), $X_{i}\cap X=\emptyset$, so contractions of the type \eqref{C1} are always finite by the definition of the R-invariant.  We must also consider the contraction between a power of $\frac{\partial^{2}\cA}{\partial\chi^{2}}$ on $X_{i}$ from the operator insertion \eqref{nabelian} and a field $\cA$ on $X$; this is easily seen to give
\begin{multline}\label{C2}
\left\la \overbrace{\frac{\partial^{2}\cA}{\partial\chi^{a}\partial\chi^{b}}|_{X_{i}}\cA|_{X}} \right\ra = \frac{\partial^{2}}{\partial\chi_{A_i}^{a}\partial\chi_{B_{i}}^{b}}[A_{i},B_{i},*,A,B] \\
=\frac{\delta^{0|2}_{ab}\left(\chi_{A_i}(B_{i}*A B) +\chi_{B_{i}}(*ABA_{i})+\chi_{A}(BA_{i}B_{i}*)+\chi_{B}(A_{i}B_{i}*A)\right)}{(A_{i}B_{i}AB)(BA_{i}B_{i}*)(A_{i}B_{i}*A)}.
\end{multline}
Once again, this quantity remains finite in the null limit due to the genericity assumption.  Replacing $X_{i}$ by $Y$ yields identical results, so we see that no contractions between operator insertions and MHV vertices can contribute to the null limit.

\subsection*{\textit{Contractions between non-adjacent $X_{i}$s}}

Identical calculations to those that lead to \eqref{C1} and \eqref{C2} allow us to compute contractions between non-adjacent operator insertions on $X_{i}$ and $X_{j}$ for $j\neq i+1,\; i-1$.  In particular, it follows that contractions between frames, or between a frame and a power of $\frac{\partial^{2}\cA}{\partial\chi^{2}}$ in such configurations are given respectively by:
\be{C3}
\left\la \overbrace{\cA|_{X_{i}}\cA|_{X_{j}}} \right\ra=\int_{\C^{2}} \frac{\d s_{i}}{s_{i}}\frac{\d s_{j}}{s_j} \Delta_{*}(Z(s_{i}), Z(s_j))=[A_{i},B_{i},*,A_{j},B_{j}],
\ee
\begin{multline}\label{C4}
\left\la \overbrace{\frac{\partial^{2}\cA}{\partial\chi^{a}\partial\chi^{b}}|_{X_{i}}\cA|_{X_{j}}} \right\ra = \frac{\partial^{2}}{\partial\chi_{A_i}^{a}\partial\chi_{B_{i}}^{b}}[A_{i},B_{i},*,A_{j},B_{j}] \\
=\frac{\delta^{0|2}_{ab}\left(\chi_{A_i}(B_{i}*A_{j}B_{j}) +\chi_{B_{i}}(*A_{j}B_{j}A_{i})+\chi_{A_{j}}(B_{j}A_{i}B_{i}*)+\chi_{B_{j}}(A_{i}B_{i}*A_{j})\right)}{(A_{i}B_{i}A_{j}B_{j})(B_{j}A_{i}B_{i}*)(A_{i}B_{i}*A_{j})}.
\end{multline}
Finally, we must compute the contraction between an operator insertion of $\frac{\partial^{2}\cA}{\partial\chi^{2}}$ on each line:
\begin{multline}\label{C5}
\left\la \overbrace{\frac{\partial^{2}\cA}{\partial\chi^{a}\partial\chi^{b}}|_{X_{i}}\frac{\partial^{2}\cA}{\partial\chi^{c}\partial\chi^{d}}|_{X_{j}}} \right\ra =\frac{\partial^{4}}{\partial\chi^{a}_{A_{i}}\partial\chi^{b}_{B_{i}}\partial\chi^{c}_{A_{j}}\partial\chi^{d}_{B_{j}}}[A_{i},B_{i},*,A_{j},B_{j}] \\
=\frac{\epsilon_{abcd}}{(A_{i}B_{i}A_{j}B_{j})}.
\end{multline}
Once again, because the $X_{i}$ only become \emph{pairwise} null separated in the limit, all three of \eqref{C3}-\eqref{C5} remain finite and cannot contribute in the null limit.

\subsection*{\textit{Contractions between adjacent $X_{i}$s}}

When computing the contractions between frames and operators on $X_{i}$ with those on $X_{i+1}$, care must be taken when discussing the null limit where these two lines intersect.  In particular, we need some regularization method for isolating the behavior of these contractions as the limit is approached.  Since we work at the level of the integrand, our regulator need not be gauge invariant, so we take the simplest mechanism possible: as $x^{2}_{i,i+1}\rightarrow 0$, assume that we can write $Z_{A_{i+1}}=Z_{B_i}+\varepsilon Z$ for some twistor $Z$.  Note that in this scheme, the numerator of the R-invariant \eqref{r-invariant} becomes:
\begin{multline*}
\delta^{0|4}\left(\varepsilon\chi_{A_{i}}(B_{i}* Z B_{i+1})+\;\mathrm{cyclic}\right)=\prod_{a=1}^{4}\left[\varepsilon\chi^{a}_{A_{i}}(B_{i}* Z B_{i+1})+ \chi^{a}_{B_i}(Z B_{i}B_{i+1}A_{i}) \right.\\
 \left. +\varepsilon\chi^{a}_{B_i}(* Z B_{i+1}A_{i})+\chi^{a}_{B_i}(B_{i+1}A_{i}B_{i}*)+\varepsilon\chi^{a}_{B_{i+1}}(A_{i}B_{i}* Z)\right] \sim O(\varepsilon^{4}),
\end{multline*}
while the denominator behaves as:
\begin{multline*}
\varepsilon^{4}(B_{i+1}A_{i}B_{i}Z)(A_{i}B_{i}* Z)(B_{i}* Z B_{i+1})(* Z B_{i+1} A_{i})(Z B_{i+1} A_{i} B_{i}) \\
+ \varepsilon^{3}(B_{i+1}A_{i}B_{i}*)(A_{i}B_{i}* Z)(B_{i}* Z B_{i+1})(* B_{i}B_{i+1}A_{i})(Z B_{i+1}A_{i}B_{i}).
\end{multline*}

Thus, we can compute the contraction between a fields $\cA$ in the frames $U_{X_i}$ and $U_{X_{i+1}}$:
\begin{equation*}
\left\la \overbrace{\cA|_{X_{i}}\cA|_{X_{i+1}}} \right\ra=\int_{\C^{2}} \frac{\d s_{i}}{s_{i}}\frac{\d s_{i+1}}{s_{i+1}} \Delta_{*}(Z(s_{i}), Z(s_{i+1}))=[A_{i},B_{i},*,A_{i+1},B_{i+1}],
\end{equation*}
which in the null limit gives
\be{C6}
\lim_{x^{2}_{i,i+1}\rightarrow 0}\left\la \overbrace{\cA|_{X_{i}}\cA|_{X_{i+1}}} \right\ra= \lim_{\varepsilon\rightarrow 0}[A_{i},B_{i},*, (B_{i}+\varepsilon Z), B_{i+1}] \sim \lim_{\varepsilon\rightarrow 0} \frac{\varepsilon^{4}}{\varepsilon^{4}+\varepsilon^{3}} =0.
\ee
We can therefore conclude that such contractions actually vanish in the null limit; this is a consequence of $\cN=4$ supersymmetry in the numerator of the R-invariant.  As for the contraction between an insertion of $\frac{\partial^{2}\cA}{\partial\chi^{2}}$ on $X_{i}$ and a field $\cA$ in $U_{X_{i+1}}$, we have
\begin{multline*}
\left\la \overbrace{\frac{\partial^{2}\cA}{\partial\chi^{a}\partial\chi^{b}}|_{X_{i}}\cA|_{X_{i+1}}} \right\ra = \frac{\partial^{2}}{\partial\chi_{A_i}^{a}\partial\chi_{B_{i}}^{b}}[A_{i},B_{i},*,A_{i+1},B_{i+1}] \\
=\frac{\delta^{0|2}_{ab}\left(\chi_{A_i}(B_{i}*A_{i+1}B_{i+1}) +\chi_{B_{i}}(*A_{i+1}B_{i+1}A_{i})+\chi_{A_{i+1}}(B_{i+1}A_{i}B_{i}*)+\chi_{B_{i+1}}(A_{i}B_{i}*A_{i+1})\right)}{(A_{i}B_{i}A_{i+1}B_{i+1})(B_{i+1}A_{i}B_{i}*)(A_{i}B_{i}*A_{i+1})}.
\end{multline*}
The null limit is then computed using our regulator
\be{C7}
\lim_{x^{2}_{i,i+1}\rightarrow 0} \left\la \overbrace{\frac{\partial^{2}\cA}{\partial\chi^{a}\partial\chi^{b}}|_{X_{i}}\cA|_{X_{i+1}}} \right\ra \sim \lim_{\varepsilon\rightarrow 0}\frac{\varepsilon^{2}}{\varepsilon^{2}} =1.
\ee
Hence, these contributions remain finite in the null limit, and also contribute nothing to the overall correlation function we are interested in calculating.

Finally, we must consider the contraction between insertions of $\frac{\partial^{2}\cA}{\partial\chi^{2}}$ on each of $X_{i}$ and $X_{i+1}$.  From \eqref{C5}, it is easy to see that
\begin{multline*}
\left\la \overbrace{\frac{\partial^{2}\cA}{\partial\chi^{a}\partial\chi^{b}}|_{X_{i}}\frac{\partial^{2}\cA}{\partial\chi^{c}\partial\chi^{d}}|_{X_{i+1}}} \right\ra =\frac{\partial^{4}}{\partial\chi^{a}_{A_{i}}\partial\chi^{b}_{B_{i}}\partial\chi^{c}_{A_{i+1}}\partial\chi^{d}_{B_{i+1}}}[A_{i},B_{i},*,A_{i+1},B_{i+1}] \\
=\frac{\epsilon_{abcd}}{(A_{i}B_{i}A_{i+1}B_{i+1})}.
\end{multline*}
Rather than use our regulator, we note that the behavior of this contraction in the null limit is evident after integrating the contraction over the respective operator insertion sites:
\begin{multline}\label{C8}
\int_{X_{i}\times X_{i+1}}\la\lambda_{i}\d\lambda_{i}\ra\la\lambda_{i+1}\d\lambda_{i+1}\ra \left\la \overbrace{\frac{\partial^{2}\cA}{\partial\chi^{a}\partial\chi^{b}}|_{X_{i}}\frac{\partial^{2}\cA}{\partial\chi^{c}\partial\chi^{d}}|_{X_{i+1}}} \right\ra\\
=\la A_{i}B_{i}\ra \la A_{i+1}B_{i+1}\ra \frac{\partial^{4}}{\partial\chi^{a}_{A_{i}}\partial\chi^{b}_{B_{i}}\partial\chi^{c}_{A_{i+1}}\partial\chi^{d}_{B_{i+1}}} \int_{\C^{2}}\frac{\d s_{i}}{s_{i}}\frac{\d s_{i+1}}{s_{i+1}} \Delta_{*}(Z(s_{i}),Z(s_{i+1})) \\
=\epsilon_{abcd}\frac{\la A_{i}B_{i}\ra \la A_{i+1}B_{i+1}\ra}{(A_{i}B_{i}A_{i+1}B_{i+1})}=\frac{\epsilon_{abcd}}{(x_{i}-x_{i+1})^{2}}.
\end{multline}
Such a contraction thus diverges as $x^{2}_{i,i+1}\rightarrow 0$ in the null limit, precisely the behavior needed to cancel the contribution from the tree-level denominator of \eqref{nplim1}.  Consequently, it is only these contractions which survive in the null limit.

\bigskip

After all such contractions have been performed and we have passed to the null limit, it is clear that all that remains in the path integral is the trace over frames $U_{X_{i}}$ and the operator in general position $\cO(y)$.  As the trace over frames is precisely the twistor Wilson loop in the adjoint representation (at the level of the integrand), the proof of proposition \ref{locP1} is complete.  The relations stated in corollary \ref{locP2} can also be checked using the contraction formulae stated in \eqref{C1}-\eqref{C8}.

\bibliographystyle{JHEP}
\bibliography{localops}

\end{document}